\documentclass{aa}  

\usepackage{graphicx}
\usepackage{txfonts}
\usepackage{color}
\newcommand{\kms}{km\,s$^{-1}$}
\newcommand{\vsini}{$v \sin i$}
\newcommand{\teff}{$T_{\rm eff}$}
\newcommand{\logg}{$\log\,{g}$}

\newcommand{\mstar}{$M_{\star}$}
\newcommand{\rstar}{$R_{\star}$}

\newcommand{\msun}{$M_{\odot}$}

\newcommand{\tic}{TIC\,$277566483$}
\newcommand{\toi}{TOI\,$1356.01$}
\newcommand{\hd}{HD\,235349}

\begin{document} 

   \title{A rare phosphorus-rich star in an eclipsing binary from TESS}
   \author{Colin P. Folsom\inst{\ref{inst1},\ref{inst2}}
          \and
          Mihkel Kama\inst{\ref{inst3},\ref{inst1}}
          \and
          T\~{o}nis Eenm\"{a}e\inst{\ref{inst1}}
          \and
          Indrek Kolka\inst{\ref{inst1}}
          \and
          Anna Aret\inst{\ref{inst1}}
          \and
          Vitalii Checha\inst{\ref{inst1}}
          \and
          Anni Kasikov\inst{\ref{inst1}}
          \and
          Laurits Leedj\"{a}rv\inst{\ref{inst1}}
          \and
          Heleri Ramler\inst{\ref{inst1}}
        }

   \institute{Tartu Observatory, University of Tartu, Observatooriumi 1, T\~{o}ravere, 61602, Estonia\\
   \email{colin.folsom@ut.ee}\label{inst1}
         \and
    Department of Physics \& Space Science, Royal Military College of Canada, PO Box 17000 Station Forces, Kingston, ON, K7K~0C6, Canada\label{inst2}
         \and
    Department of Physics and Astronomy, University College London, Gower Street, London, WC1E 6BT, UK\label{inst3}
             }

 
  \abstract
   {Few exoplanets around hot stars with radiative envelopes have been discovered, although new observations from the TESS mission are improving this. Stars with radiative envelopes have little mixing at their surface, and thus their surface abundances provide a sensitive test case for a variety of processes including potentially star-planet interactions. Atomic diffusion is particularly important in these envelopes, producing chemically peculiar objects such as Am and HgMn stars.
   }
   {An exoplanet candidate around the B6 star HD 235349 was identified by TESS.  Here we determine the nature of this transiting object and identify possible chemical peculiarities in the star.}
   {HD 235349 was observed using the long-slit spectrograph at Tartu Observatory, as well as photometrically by the TESS mission.  The spectra were modeled to determine stellar parameters and chemical abundances.  The photometric light curve was then analyzed in the context of the stellar parameters to determine properties of the transiting object.}
   {We find the transiting object is a low-mass stellar companion, not a planet.  However, the primary of this eclipsing binary is a rare type of chemically peculiar star. A strong overabundance of P is found with overabundances of Ne and Nd, and mild overabundances of Ti and Mn, while He is mildly underabundant.  There is also clear evidence for vertical stratification of P in the atmosphere of the star. The lack of Hg and weak Mn overabundance suggests that this is not a typical HgMn star. It may be in the class of helium-weak phosphorus-gallium (He-weak PGa) stars, or an intermediate between these two classes.}
   {We show that \hd\ is a rare type of chemically peculiar star (He-weak PGa) in an eclipsing binary system with a low-mass stellar companion.  This appears to be the first He-weak PGa star discovered in an eclipsing binary.} 


   \maketitle
%
\section{Introduction}\label{sec:intro}

The Transiting Exoplanet Survey Satellite (TESS) mission \citep{TESS2015}, recently identified \object{\hd}\ (\object{\toi}, \tic, $\alpha_{J2000}=20^{h}44^{m}46^{s}.7$, $\delta_{J2000}=+54^{\circ}30'07''.86$) as an Object of Interest based on photometric transits.
The TESS lightcurve of \hd\ contains a periodic transit signature with a depth of $0.75\,$ per cent and a period of $24.28546 \pm 0.00102$ days, reported in the Exoplanet Follow-up Observing Program for TESS (ExoFOP-TESS)\footnote{\texttt{https://exofop.ipac.caltech.edu/tess/} } database (and later confirmed by our own inspection of transit times).  Based on discrepant previous effective temperature values the star may be among the hottest exoplanet hosts to-date or, alternatively, the large primary of an eclipsing binary. In order to clarify its nature, we carried out a comprehensive analysis of \hd\ including a detailed chemical composition determination and a radial velocity time series. 

\hd\ is an early-type star with a Gaia\,Early Data Release 3 (EDR3) parallax distance of $d = 1901 \pm 180\,$pc \citep{Bailer-Jones2021}. It was chosen as one of the first targets in the PIVOT (Planetary InitatiVe at the Observatory of Tartu) project, which is aimed at characterizing stars with exoplanets, protoplanetary disks, or chemical peculiarities. The TESS Object of Interest catalog lists a transit depth of $0.75\,$ per cent and a classification PC (planetary candidate). The two different primary star temperatures listed by the ExoFOP-TESS database place the transiting companion as either a giant planet (if \teff\,$=8381\pm409\,$K) or a low-mass star (if \teff\,$=12\,308\pm144\,$K). These and other available \teff\ estimates for the primary disagree at high confidence, further motivating a follow-up analysis.

Stars with spectral type earlier than F5 (\mstar\,$ \geq 1.4\,M_\sun$) are of particular interest because, in addition to the more familiar planet-metallicity correlation \citep{Gonzalez1997, FischerValenti2005}, causal links may exist between certain chemically peculiar early-type stars and their planetary systems \citep{Jura2015, Kamaetal2015}. More generally, due to the lack of a convective envelope, superficial chemical peculiarities arise in early-type stars for a variety of reasons, such as radiative levitation \citep[Am stars;][and references therein]{10.1093/mnras/stz080}, magnetic flotation with slow rotation \citep[Ap/Bp;][]{2020pase.conf...35M}, gravitational settling with slow rotation and lack of magnetic fields \citep[HgMn;][]{Alecian-Michaud1981, Makaganiuk2011-HgMn-nonMag}, or ``accretion contamination'' due to a lack of strong mixing \citep[e.g., $\lambda\,$Bo\"{o};][]{Paunzen1998}.

We describe our spectroscopic observations in Sect.\,\ref{sec:obs}. In Sect.\,\ref{sec:analysis}, we detail the spectral fitting, abundance analysis, and the radial velocity and photometric lightcurves.  We present the resulting stellar composition and parameters in Sect.\,\ref{sec:results}, as well as constraints on the secondary's properties. The chemical peculiarity of \hd\ and the nature of its secondary are discussed in Sect.\,\ref{sec:discussion}.

\section{Observations}\label{sec:obs}

Spectroscopic observations of \hd\ were carried out at Tartu Observatory\footnote{\texttt{https://kosmos.ut.ee/en}} (58$^\circ$15'57''\,N,  26$^\circ$27'35''\,E) in Estonia with the 1.5-meter Cassegrain reflector AZT-12 (see Appendix\,\ref{AppendixA} for full details) on eleven nights from June, 2020 to April, 2021. A typical session included observations of radial velocity and spectrophotometric standard stars. The long-slit spectrograph ASP-32 at the Cassegrain focus was used with 1800\,lines/mm and 1200\,lines/mm gratings. The wavelength coverage extended from $3704\,$\AA\ to $7943\,$\AA\ with signal-to-noise ratios (S/N) from $\approx 70$ to $200$. The full list of observations is given in Table\,\ref{tab:observations}.

The data were processed using the {\sc iraf}\footnote{\texttt{https://iraf.net}} \citep{Tody1986} software built-in packages for CCD pre-processing, spectroscopy, and cross-correlation. Spectrophotometric standards (Vega and 10\,Lac) from the CALSPEC database \citep{2020AJ....160...21B} were used to correct for the complex instrumental profiles of some observed spectra, but absolute flux calibrations were not attempted. After bias removal and flat fielding with a lamp, spectral extraction was performed using the long-slit spectrum processing routines in \texttt{ctioslit} and \texttt{onedspec} packages. Sigma spectra were used to estimate the S/N of science spectra. Thorium-argon arc lamp spectra were recorded before and after each science integration, their dispersion solutions were averaged. Heliocentric correction was carried out with the {\sc iraf} tasks \texttt{rvcorrect} and \texttt{dopcor}. For abundance analysis, spectra were averaged nightly. The continuum was normalized by fitting a low degree cubic spline using a custom program created in Python programming language. The spectral resolution was measured using Gaussian fits to well-isolated ThAr emission lines close to the central wavelength in each spectral setup. The average Gaussian full width at half maximum was used as the width of instrumental profile of the spectrograph.

\begin{table}[htb]
\caption{Spectra of \hd\ obtained with the 1.5~m telescope at Tartu Observatory.}
\centering
\begin{tabular}{ccccccc@{}}
\hline\hline
Date & $\lambda_\mathrm{min}$ & $\lambda_\mathrm{max}$ & $\lambda/\Delta\lambda$ & Exp. & S/N & grating\\
 & (\AA) & (\AA) & & (s) & & l/mm\\
\hline
2020-06-18 & 4795 & 5666 & 2870 & 1800 & 130 & 1200\\
2020-06-18 & 5517 & 6329 & 4210 & 1800 & 150 & 1200\\
2020-06-18 & 6228 & 6978 & 4500 & 1800 & 140 & 1200\\
2020-09-19 & 3803 & 4750 & 2620 & 2700 & 70  & 1200\\
2020-09-19 & 3993 & 4509 & 5030 & 4500 & 130 & 1800\\
2020-09-19 & 4344 & 4827 & 5560 & 5400 & 130 & 1800\\
2020-09-19 & 4887 & 5320 & 6490 & 5400 & 200 & 1800\\
2020-09-19 & 5278 & 5672 & 7200 & 3600 & 150 & 1800\\
2020-11-20 & 5915 & 6241 & 8646 & 3600 & 100 & 1800\\
2020-11-23 & 5915 & 6240 & 8646 & 4800 & 140 & 1800\\
2021-03-22 & 4864 & 5298 & 6910 & 5400 & 140 & 1800\\
2021-04-01 & 4889 & 5321 & 6760 & 5400 & 104 & 1800\\
2021-04-06 & 4889 & 5321 & 6830 & 5400 & 150 & 1800\\
2021-04-10 & 4889 & 5321 & 6780 & 5400 & 140 & 1800\\
2021-04-18 & 3704 & 4244 & 4840 & 10800& 90  & 1800\\
2021-04-19 & 4889 & 5321 & 6750 & 5400 & 140 & 1800\\ 
\hline
\end{tabular}
\label{tab:observations}
\end{table}

\section{Analysis}\label{sec:analysis}

\subsection{Fundamental parameters based on spectroscopy}

\subsubsection{Balmer line analysis}
\label{sec:BalmerLineAnalysis}

Since the fundamental parameters of \hd\ are not well constrained in the literature, we attempted to derive them from our spectroscopic observations.  Balmer lines can provide strong constraints on \teff\ and \logg.  The observations show no sign of emission lines or infilling, and the star is cool enough that the stellar wind should not contribute substantially to the observed lines.  Thus we proceeded by directly fitting model spectra to the observations.

To calculate the model spectra we used the {\sc zeeman} spectrum synthesis code \citep{Landstreet1988, Wade2001}, and {\sc atlas9} model atmospheres \citep{Kurucz1993-ATLAS9etc,CastelliKurucz2003}.  Atomic line data are taken from the Vienna Atomic Line Database \citep[VALD,][]{Piskunov1995-VALD, Ryabchikova1997-VALD, Kupka1999-VALD2, Ryabchikova2015-VALD3}.  

{\sc zeeman} has been mostly used for metallic line analysis, particularly of magnetic and chemically peculiar stars. Since it is preferable to use one code for all the spectrum synthesis, we have added the calculation of hydrogen line profiles to the code. 
The hydrogen Stark broadening profiles use the grid calculated by \citet{Lemke1997}, which implements the widely used VCS model of \citet{Vidal1973}.  Stark broadening profiles are interpolated as necessary, and are implemented for the Lyman through Brackett series, up to $n=22$ of each series. 
Resonance self broadening is calculated using the theory of \citet{AliGriem1965,AliGriem1966}, and transition data for this calculation were taken from the National Institute of Standards and Technology (NIST) Atomic Spectra Database \citep{NIST_ASD}. 
More advanced calculations exist, and may be implemented in the future, but for the current analysis self broadening is much weaker than Stark broadening.
The broadening calculations also include thermal Doppler broadening, van der Waals broadening (with coefficients from VALD, although it is generally negligible), and microturbulence.  In our implementation we use the tabulated Stark broadening profiles including thermal broadening of \citet{Lemke1997}, which is convolved with a Voigt profile that includes the other broadening terms, and then interpolated onto the wavelength grid used for the model spectrum.   
The model hydrogen lines here do not include the Zeeman effect, and thus are likely insufficient for strongly magnetic Ap/Bp stars, or polarized spectra, but should be sufficient for \hd.

To verify the accuracy of the Balmer line calculations, we tested the spectra against the Kurucz {\sc balmer} \citep{Kurucz1993-ATLAS9etc} and {\sc synthe} codes (\citealt{Kurucz1993-ATLAS9etc}, with grids of model spectra from \citealt{Munari2005-SYNTHE-grid} and \citealt{Bertone2008-SYNTHE-grid}), which have similar physics to {\sc zeeman}.  This produces an excellent agreement over a wide range of temperatures, provided the same model atmospheres are used. 
Testing against the {\sc phoenix} code for K-type stars (\citealt{Allard2012}) provides very good agreement.  {\sc phoenix} models for hotter stars (\citealt{Husser2013}, up to 12000 K) provide an acceptable agreement, although these models have slightly narrower wings, which in the worst case could lead to a discrepancy in the inferred \logg\ up to 0.1. 
Testing against a grid of models from {\sc pfant} \citep{Coelho2005} provides a very good agreement, similar to {\sc synthe}, although this grid only extends to 8000 K.
Testing against spectra from the {\sc bstar2006} grid of the {\sc tlusty} code (for cooler B-type stars in the grid, \citealt{Lanz2007-BSTAR2006}) provides excellent agreement, except for the line cores of H$\alpha$ and H$\beta$.  

In observations of Balmer lines, a very precise continuum normalization is difficult and sometimes not practically possible.  To account for possible errors in the continuum placement of the observation, we include a continuum polynomial in the model spectrum for fitting Balmer lines.  For this we use a simple quadratic polynomial, in the form $c_0 + c_1 \lambda + c_2 \lambda^2$. Including this significantly reduced the scatter in best fit \teff\ and \logg\ from individual Balmer lines. 

In order to determine optimal \teff\ and \logg\ values, and probe their probability distributions, we adopted a Markov chain Monte Carlo (MCMC) approach.  This is more computationally intensive than simply fitting by $\chi^2$ minimization, but offers two advantages.  First, there tends to be a covariance between \teff\ and \logg, and MCMC allow us to characterize the correlation in their formal uncertainties.  Second, while we include three coefficients for our continuum polynomial, and they will have some influence on the uncertainties of \teff\ and \logg, we do not care about the values of those coefficients, they are effectively nuisance parameters.  An MCMC approach allows us to marginalize over these parameters (in a Bayesian sense), to derive distributions of the parameters of interest.
We used the {\sc  emcee} package from \citet{Foreman-Mackey2013}, which uses the MCMC affine-invariant ensemble sampler of \citet{GoodmanWeare2010}.  This package has the advantages of being robustly tested and easy to integrate with existing code.  The integration with {\sc emcee} was implemented as a Python wrapper around the existing Fortran code of {\sc zeeman}.  

For this analysis we used the lower resolution observation from 2020-09-19 with $\lambda/\Delta\lambda = 2620$, spanning 3803--4750 \AA, since the resolution is sufficient for the Balmer lines and this provides H$\gamma$ through H9 in one observation.  Consistent results were obtained from the higher resolution observation from the same night spanning 3933--4509 \AA, covering the H$\gamma$ and H$\delta$ lines.  
The results for individual lines are presented in Appendix \ref{AppendixTeff} and Fig.~\ref{fig:balmerLine-cornerPlots}.
To get the final values shown in Table \ref{tab:abundances}, we take the median of the \teff\ and \logg\ distributions for each line, and take the average over the 5 lines as the global best value and the standard deviation as the formal uncertainty.  This standard deviation is consistent with the uncertainties on individual lines. 

\subsubsection{Helium and metal line analysis}
\label{sec:MetalLineAnalysis}

In order to derive chemical abundances for a range of elements, as well as \vsini\ and microturbulence ($v_{\rm mic}$), we fit observations of metallic lines across most of the visible range.  
A secondary analysis where we derive \teff\ and \logg\ from metal lines simultaneously with chemical abundances is discussed in Appendix \ref{AppendixTeff}.
Synthetic spectra were calculated with {\sc zeeman}, using {\sc atlas9} model atmospheres and atomic line data from VALD. The VALD line lists were obtained from an `extract stellar' request with enhanced abundances for peculiar elements.  The hyperfine splitting (HFS) calculations in VALD from \citet{Pakhomov-VALD-HFS} were included in the line list.  Testing with and without hyperfine splitting components produced consistent results for the best fit abundances, well within our uncertainties.\footnote{
The impact of HFS on our final results is small mostly because, at this \vsini\ and resolution, HFS largely has the effect of desaturating strong lines.  Lines of elements with a significant nuclear dipole moment (and HFS data in VALD) are relatively weak in this spectrum, so desaturation due to HFS is very small (generally less than 1\% of our uncertainties on individual pixels).  However, the effect is present in lines of interest to this study (particularly Mn {\sc ii} and Ga {\sc ii}), thus it should be included, and may become significant for stars with stronger or sharper lines. }

Stellar parameters were derived by directly fitting synthetic spectra to the observations using the $\chi^2$ minimization code of \citet{Folsom2012-HAeBe-abun}.  Since \ion{He}{I}  lines are prominent in the spectrum and in the He abundance is important, we included the Stark broadened profiles of \citet{Barnard1969, Barnard1974, Barnard1975} in the opacity profile calculations for the \ion{He}{I} 4471, 4387, 4026, and 4921 \AA\ lines.  

\begin{figure*}[!ht]
    \centering
    \includegraphics[width=1.0\linewidth]{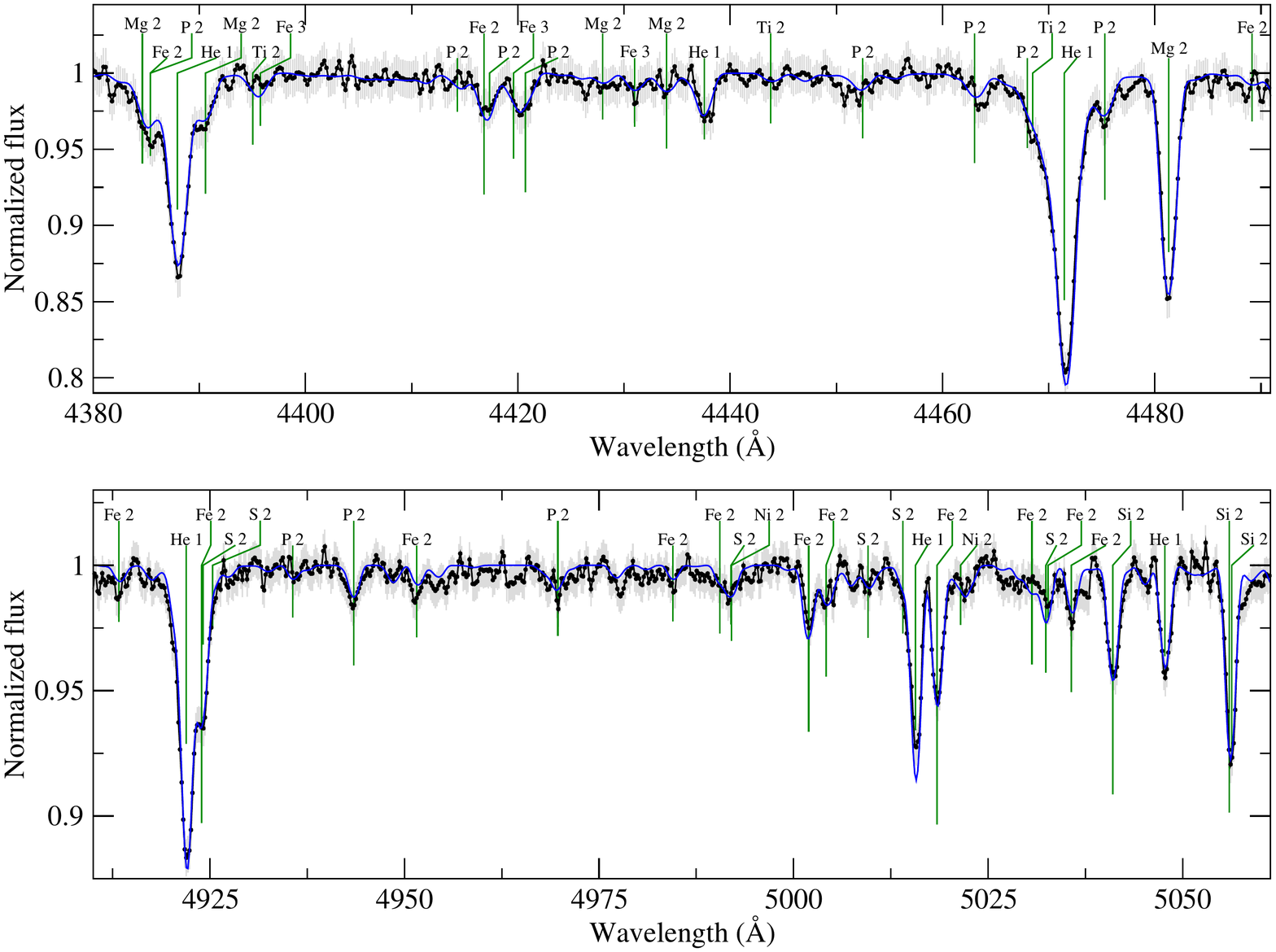}
    \caption{Observed spectrum of HD 235349 obtained with the Tartu Observatory long-slit
spectrograph (black, error bars light gray), and the best fit model spectrum (blue). Species contributing to the stronger lines in the spectrum are indicated (green). }
    \label{fig:spectrum1}
\end{figure*}

For this analysis we used the higher resolution observations (obtained with the 1800 l/mm grating).  While the spectral resolution is not particularly high, it is still well below the rotational broadening of the star, so it is sufficient for an abundance analysis.  The combination of these observations cover $\sim$330 to $\sim$500 \AA, and generally contain a good number of lines for simultaneously determining stellar parameters, thus we fit each observation independently.   Where we had multiple observations covering the same wavelength range we used the highest S/N observation.  
(The 7292--7943 \AA\ observations were not used, since they contained few detectable lines of interest.)  
This provides us with 5 spectral windows covering 4014--4500, 4375--4742, 4910--5310, 5281--5663, and 6007--6229 \AA. Balmer lines and telluric lines were excluded from these regions.  While Balmer lines could be fit simultaneously with the metal lines, they would dominate the resulting $\chi^2$ (and constraints on \teff\ and \logg), thus we prefer to fit them separately so as to obtain an independent constraint from metal lines.  Sample best fit spectra are shown in Fig.~\ref{fig:spectrum1}.

When comparing our best fit models to the observations, we noted a number of \ion{S}{II} lines with discrepant line strengths.  Comparing oscillator strengths for these lines from the VALD and NIST databases, we find significant differences in the theoretical value, with the NIST values rated `C' or better.  Using the NIST $\log gf$ values produced more consistent results, improving the fit to the observations, thus we adopted the NIST $\log gf$ values for \ion{S}{II} when available.  For similar reasons we adopted the NIST $\log gf$ values for the  \ion{P}{II} 5152.2 and 5191.4 \AA\ lines and \ion{Si}{II} 5041.0, 5055.98, and 5056.3 \AA\ lines.  

\begin{figure}[!ht]
    \centering
    \includegraphics[width=1.0\linewidth]{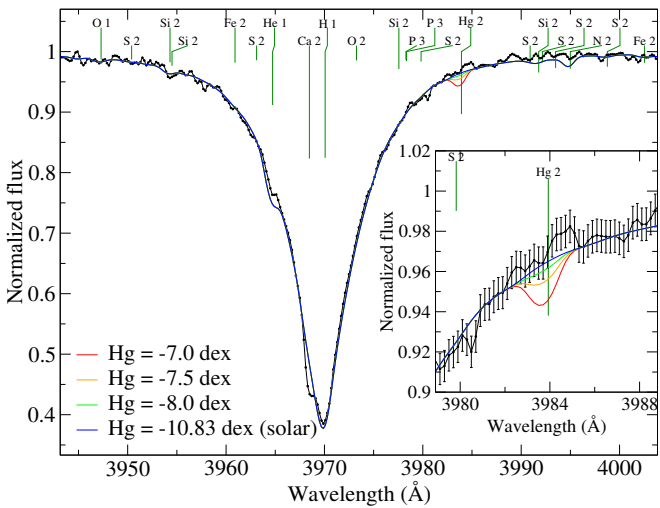}
    \caption{Observed spectrum of HD 235349 (black), and model spectra varying the Hg abundance.  H$\epsilon$ is well fit, but the Hg 3983.9 \AA\ line is not detected.
    }
    \label{fig:spectrumHg}
\end{figure}

For each of the five identified spectral regions above, we fit for \vsini, $v_{\rm mic}$, and chemical abundances for any elements with detected lines. 
The 4014--4500 and 4375--4742 regions did not produce reliable values of $v_{\rm mic}$, thus for these regions we assumed the average value from the other three regions.
For the elements Cr, Ga, Pr, and Hg, we do not detect any individual lines, however they are of interest for determining the type of chemical peculiarity in a star, particularly for HgMn stars and helium-weak phosphorus-gallium (He-weak PGa) stars.  Thus we used the strongest theoretically predicted line in our observed ranges to place upper limits on the elements.  These limits were determined by eye, taking the largest abundance that produced a model reasonably consistent with the observation. The Hg line at 3983.9~\AA\ is particularly interesting for this, as it is usually detected in HgMn stars, and its absence in \hd\ (illustrated in Fig. \ref{fig:spectrumHg}) suggests this is not a typical HgMn star.

To produce the final results we take the average of the values for each spectral region, and as an uncertainty we use the standard deviation.  
For elements with abundances from less than three regions, the uncertainty was estimated by eye, accounting for the scatter between different lines, the impact of blending lines, noise in the observation, and potential normalization errors.  For elements with abundances from three or more regions we verified that the standard deviation was reasonably consistent with the scatter between lines within one region and the noise in the observations.  These uncertainties, from the scatter in independent fits to different spectral regions, should account for errors in the atomic line data and normalization, however they should still be considered somewhat approximate.  

\subsubsection{Stratification of phosphorus}
\label{sec:AnalysisStratification}

In the initial abundance analysis, we found a clear wavelength dependence in the P abundance from different spectral windows, while the other abundances and stellar parameters had no significant correlation with wavelength.  This could be a hint of vertical stratification in abundance of that element, since the excitation potential of the lower level, and implicitly the depth that the line is formed at, also correlate with wavelength.  To investigate this we first looked at chemical abundances derived from individual lines, and then considered a simple parametric model to improve the global fit to all lines.

\begin{figure*}
    \centering
    \includegraphics[width=0.8\linewidth]{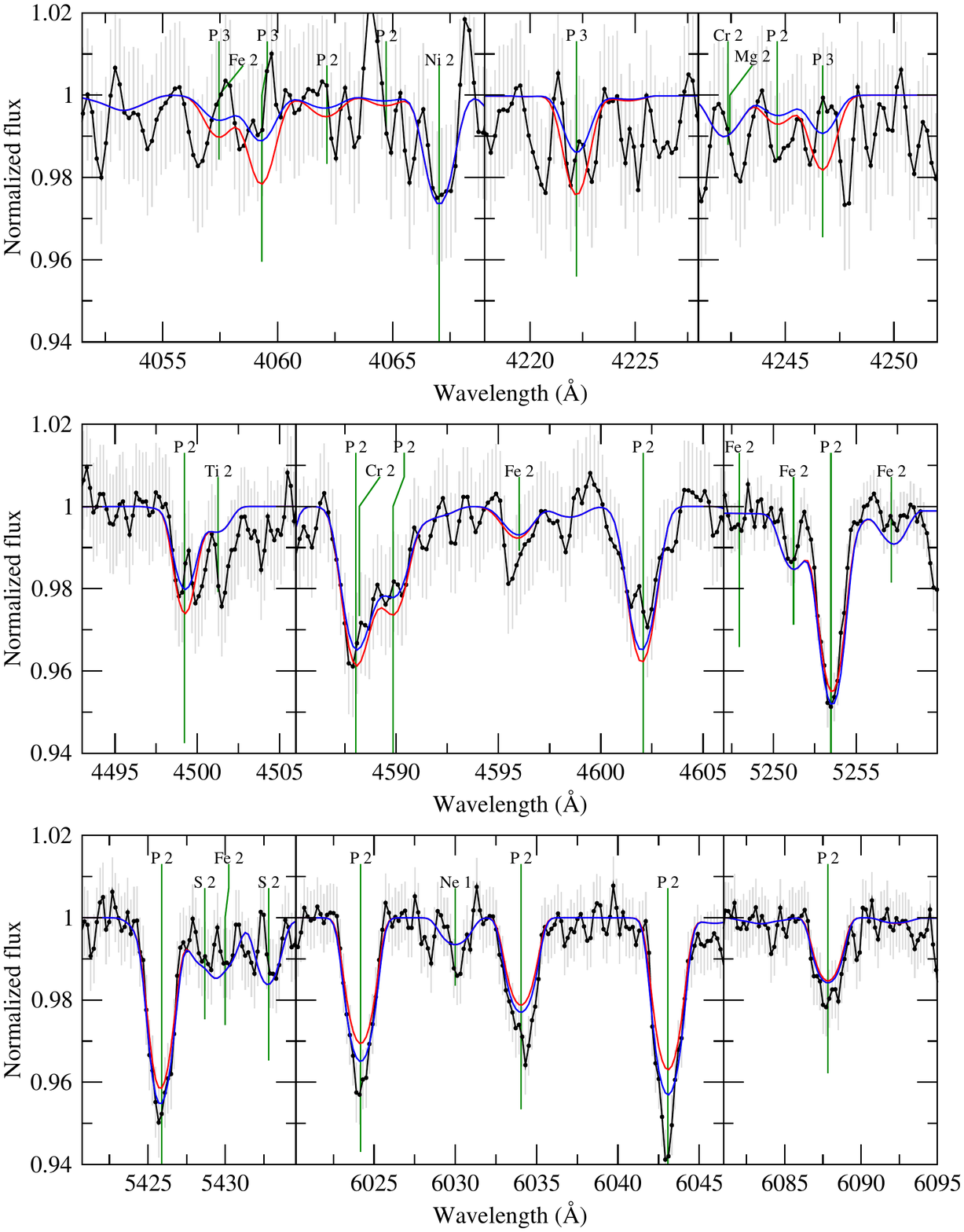}
    \caption{A selection of the P lines used for the stratification analysis.  The observation is shown in black (error bars in light gray), the best fit model with a vertically constant abundance is in red, and the best fit model with a linear variation in the abundance with log optical depth is in blue.  The stratified model does a better job of reproducing the observations than the constant model, but higher S/N observations could provide stronger constraints.
    }
    \label{fig:sample-P-lines}
\end{figure*}

\begin{figure}
    \centering
    \includegraphics[width=\linewidth]{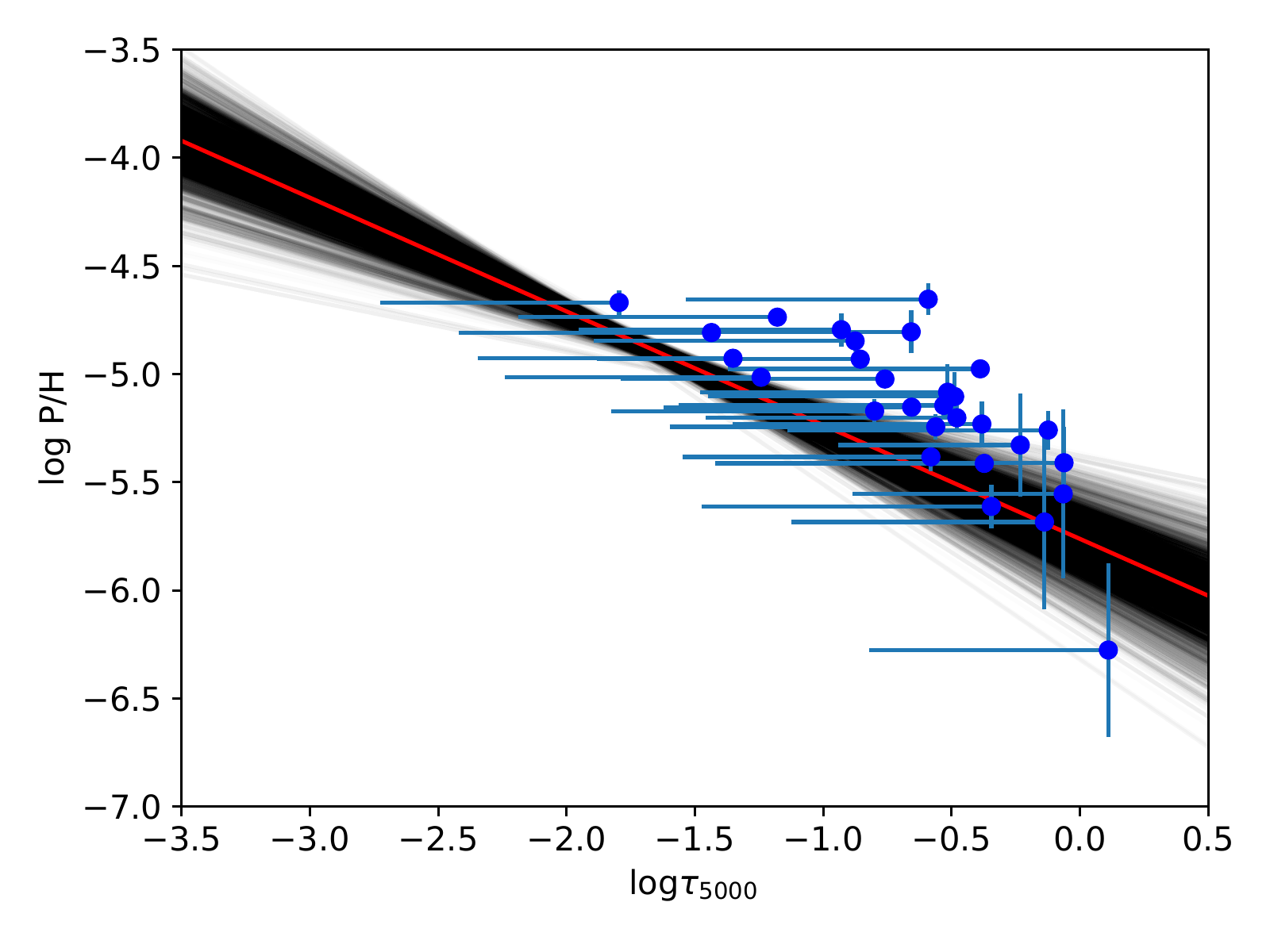}
    \caption{Abundance estimates of P as a function of optical depth ($\tau_{5000}$).  Abundances for individual lines are shown as points where the optical depth in the line center reaches unity.  The horizontal bar extends to where the optical depth in the line center is 0.1, to illustrate the range of depths relevant for line formation.  Vertical errors are the statistical uncertainty on the abundance. The red line is the best (median) linear P distribution from directly fitting all observed P lines simultaneously.  The light gray lines represent individual models from the Markov chain, with darker gray areas corresponding to a higher probability density.  }
    \label{fig:strat-P-abun}
\end{figure}

In order to derive best fit chemical abundances for individual P lines, we first adopt the best fit stellar parameters and abundances for other elements from Table \ref{tab:abundances}.  We then look for P lines that are at least marginally detected above the noise and are not in a blend that is dominated by a different element.  This produced 29 usable lines between 4045 and 6089 \AA.  Several lines of varying quality are illustrated in Fig.\ref{fig:sample-P-lines}.  We used the procedure of \citet{Khalack2007-stratification} to derive a proxy for the depth that each line is formed at.  This takes the point in the atmosphere where the optical depth at the line center ($\tau_\ell$) reaches 1. That depth is reported in terms of the optical depth in the continuum at 5000 \AA\ ($\tau_{5000}$).  Real lines form over a range of depths in the star, with important contributions from above $\tau_\ell = 1$, however this proxy is useful when investigating chemical stratification and looking for correlations between abundance and optical depth.  To approximately account for this range of depths, we also calculated the point where $\tau_\ell = 0.1$.  We plot the abundance found for each line against the optical depth for that line in Fig.~\ref{fig:strat-P-abun}, and find a clear correlation with larger abundances for lower optical depths.  

As a second approach we fit a simple parametric model for the vertical distribution of P, by calculating model spectra with this distribution and fitting them to the 29 lines identified above simultaneously.  This approach requires an assumption about the functional form of the P distribution, and we considered three options.  The first was a linear relation between the P abundance and $\log \tau_{5000}$, the second was a step function with two constant abundances above and below a transition at one depth, and the third was a linear (in $\log \tau_{5000}$) transition region with constant abundances above and below.  All three options reached nearly identical, statistically consistent, $\chi^2$ values for the best fit models.  However the second and third option require more free parameters, and those parameters became increasingly poorly constrained.  Thus we conclude that we cannot place a strong constraint on the functional form of the stratified distribution, likely since many of the P lines are very near the noise level in our observations, and adopt the simplest function: a linear relation.  

To explore the probability distribution of the parameters, and potentially strong co-variances, we used the same MCMC routine as was used for the Balmer lines in Sect.~\ref{sec:BalmerLineAnalysis}.  The free parameters of the model are the abundance at $\tau_{5000} = 1$ and the slope of abundance ($\log \rm P/H$) with $\log \tau_{5000}$.  The results of modeling these lines are presented in Sect. \ref{sec:ResultsAbundances}, with the best fit model in Fig.~\ref{fig:sample-P-lines}, showing clear support for a stratified distribution of P.

\subsection{Radial velocity perturbations}

\label{sec:rvcurve}

Our spectral observations sampled 11 individual nights over a baseline of ${\approx}10\,$months. This allowed us to determine the radial velocity amplitude $K_{\rm RV}$ of the primary and constrain the nature of the transiting companion.
Radial velocity (RV) measurements were performed on the long-slit spectra using a cross-correlation approach and the {\sc iraf} task \texttt{fxcor} \citep{1994ASPC...61...79F}.  
The correlation template was a normalized $R=20\,000$ model spectrum from \citet{Munari2005-SYNTHE-grid}, specifically the model in their grid closest to our derived parameters.
The resulting instantaneous RV values range from $-28.5$~to $7.2\,$km\,s$^{-1}$ and are listed in Table\,\ref{tab:rv}. Fitting the RV curve with a pure sine wave and assuming zero eccentricity, we find $K_{\rm RV}=(13.72\pm0.63)\,$km\,s$^{-1}$.

\subsection{Photometric lightcurve}

TESS has observed \hd\ (\toi) in sectors 15 and 16, where two exoplanet transit-like events were discovered. The depth of those events is $\approx7.5$ ppt. The photometric data from the TESS data portal has strong systematic effects, therefore we measured target and several surrounding comparison stars from cuts of TESS calibrated full-field images using traditional aperture photometry methods. Remaining residuals in the lightcurve were removed by fitting with a low order cubic spline.

\begin{figure}[htb]
    \centering
    \includegraphics[width=1.0\linewidth]{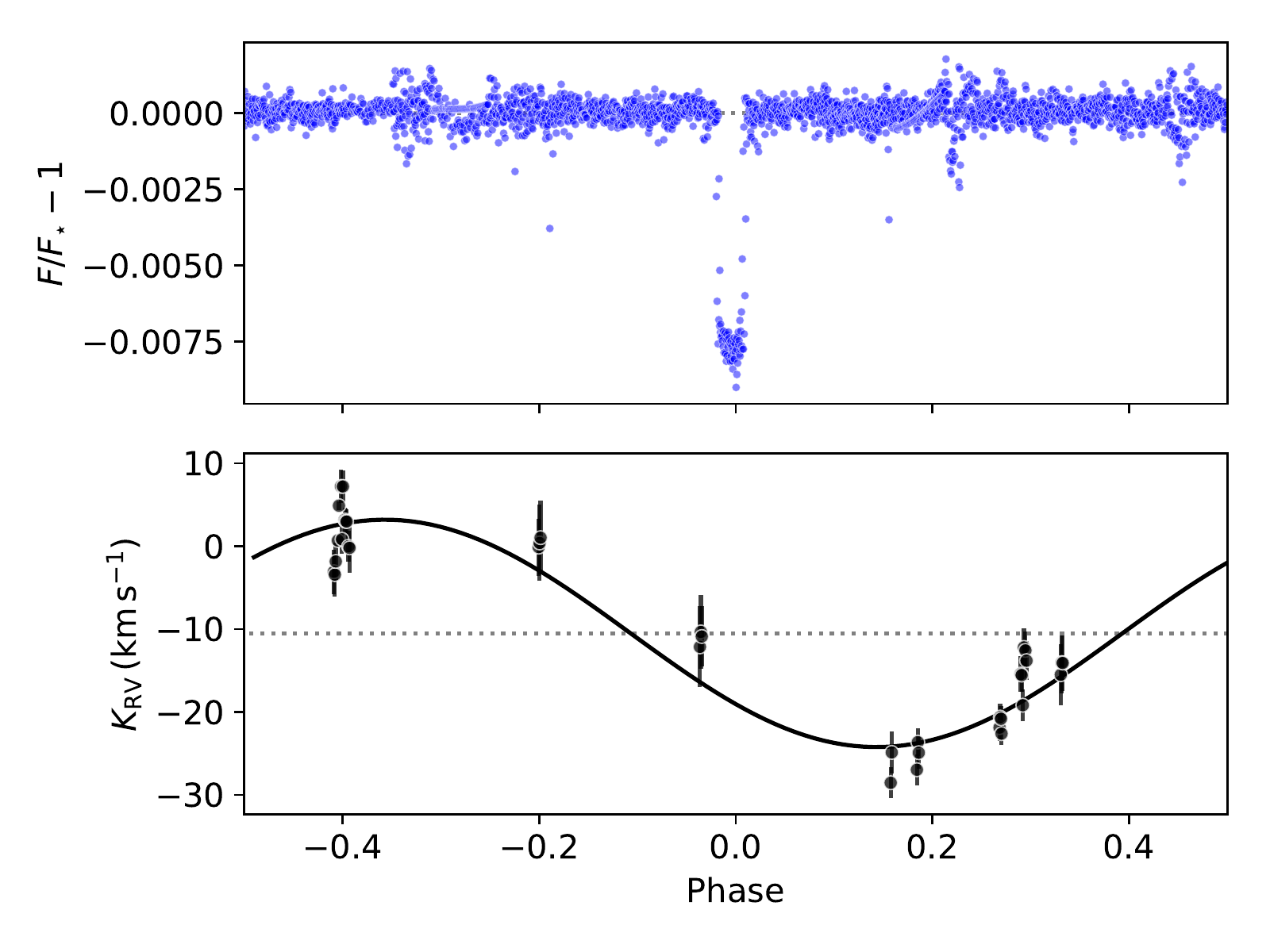}
    \caption{\emph{Upper panel: }The phase-folded TESS lightcurve for \hd. \emph{Lower panel: } The Tartu Observatory radial velocity time-series (markers) overlaid with a sinusoid fit (solid line).}
    \label{fig:phasecurve}
\end{figure}

The ExoFOP-TESS database reports a period of  $24.28546 \pm 0.00102$ days for these transits.  Phase folding our extraction of the TESS photometry with this period, we find a good agreement in the timing of the eclipses, thus we confirm the period reported in ExoFOP-TESS.
Figure~\ref{fig:phasecurve} shows the phase-folded TESS lightcurve and our radial velocity time-series for \hd. The offset of the radial velocity curve from the transit phase curve suggests an eccentric orbit, though we caution that due to possible systematics in our long-slit spectra, stable higher S/N follow-up is necessary to test this.

\section{Results}\label{sec:results}

\begin{table}[!ht]
    \centering
    \caption{Best fit stellar parameters and chemical abundances for \hd.} 
    \begin{tabular}{lccl}
\hline\hline
Parameter & Value & Solar & \#  \\  
\hline
\multicolumn{3}{l}{\emph{Balmer line averages}}\\
\teff\ (K) & $14\,757 \pm 264$ \\
\logg      & $3.34 \pm 0.11$ \\
\multicolumn{3}{l}{\emph{metal line averages}}\\
\vsini\ (\kms)       & $65.16 \pm 7.2$\\
$v_{\rm mic}$ (\kms) & $ 2.67 \pm 0.38$ \\
\hline
He   & $-1.32 \pm 0.10$ &  -1.07  & 3 \\
C    & $-3.78 \pm 0.15$ &  -3.57  & 1 \\
O    & $-3.36 \pm 0.11$ &  -3.31  & 2 \\
Ne   & $-3.42 \pm 0.23$ &  -4.07  & 2 \\
Mg   & $-4.61 \pm 0.16$ &  -4.40  & 4 \\
Al   & $-5.97 \pm 0.35$ &  -5.55  & 3 \\
Si   & $-4.83 \pm 0.16$ &  -4.49  & 5 \\
P    & $-5.11 \pm 0.20$ &  -6.59  & 5 \\
S    & $-5.35 \pm 0.22$ &  -4.88  & 4 \\
Ti   & $-6.39 \pm 0.20$ &  -7.05  & 2 \\
Cr   & $-6.3          $ &  -6.36  & 1 \\
Mn   & $-6.03 \pm 0.27$ &  -6.57  & 2 \\
Fe   & $-4.36 \pm 0.23$ &  -4.50  & 4 \\
Ni   & $-5.40 \pm 0.14$ &  -5.78  & 2 \\
Ga   & $< -6.9        $ &  -8.96  & 1 \\
Pr   & $< -8.8        $ &  -11.28 & 1 \\
Nd   & $-9.14 \pm 0.40$ &  -10.58 & 1 \\
Hg   & $< -7.5        $ &  -10.83 & 1 \\
\hline\hline
    \end{tabular}
    \tablefoot{Solar abundances from \citet{Asplund2009} are given for comparison. The number of individual spectral windows used to constrain a given element is indicated in the last column.}
    \label{tab:abundances}
\end{table}

\subsection{Stellar parameters}

\begin{figure*}
    \centering
    \includegraphics[clip=,width=0.9\linewidth]{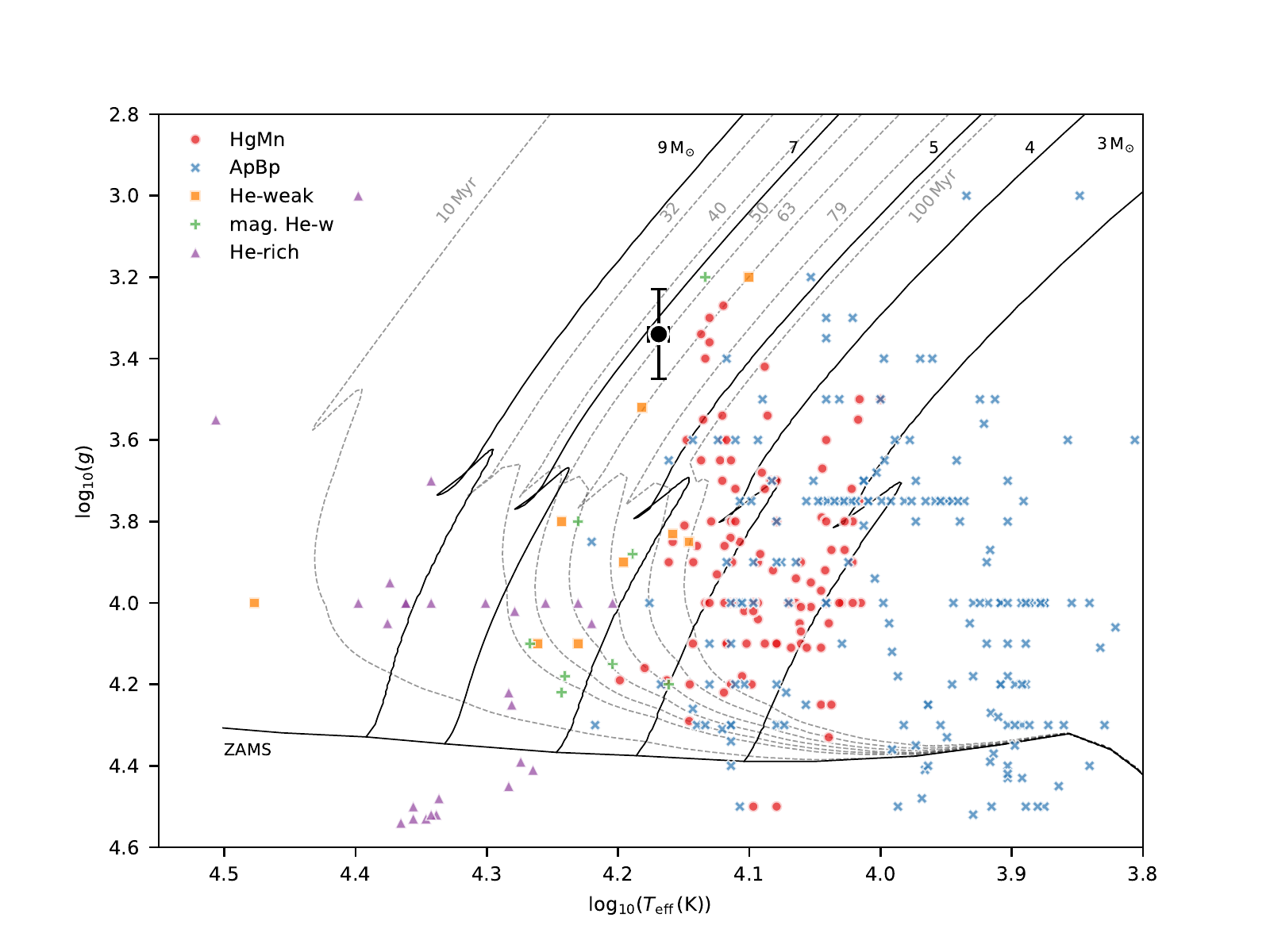}
    \caption{Evolution tracks (solid lines) and isochrones (dashed lines, solid line for ZAMS) for non-rotating stars at $Z=0.014$ from \citet{Ekstrom_etal2012A&A...537A.146E}. Annotations give the mass at ZAMS for evolution tracks and the age for isochrones. The position of \hd\ is given by the larger black point.  A comparison sample of other chemically peculiar stars from \citet{Ghazaryan2018-catalogue-Ap-Am-HgMn, Ghazaryan2019-HeWeak-HeStrong-roAp} are indicated with smaller colored points. }
    \label{fig:HRD}
\end{figure*}

The spectroscopically determined best-fit parameters of \hd\ are summarized in Table\,\ref{tab:abundances}. We adopt here the values determined from the hydrogen Balmer lines: \teff\,$= 14\,757 \pm 264\,$K and \logg\,$ = 3.34 \pm 0.11$.

The current mass and age of \hd\ were determined by comparing its position in an effective temperature ({\teff}) -- surface gravity ({\logg}) diagram with stellar evolution models without rotation at solar metallicity \citep[$Z = 0.014$;][]{Ekstrom_etal2012A&A...537A.146E}, as shown in Fig.\,\ref{fig:HRD}, and interpolating between models in log mass. From this, we find a stellar mass \mstar\,$ = 6.7^{+0.9}_{-0.8}\,M_\sun$, radius \rstar\,$= 9.2^{+1.8}_{-1.5}\,R_\sun$, and luminosity $\log (L_{\star}/L_\sun)= 3.55 \pm 0.19$.  
We adopt these as our final values. As a consistency check, empirical calibrations of \mstar\ and \rstar\ as polynomial functions of \teff, {\logg} and [Fe/H] \citep{2010A&ARv..18...67T} give consistent results with the above numbers.

The distance to \hd\ was recently determined to be $d = 1901 \pm 180\,$pc using the Gaia\,EDR3 parallax \citep[][]{Bailer-Jones2021}. 
However, the current value may not be entirely reliable, since the Renormalised Unit Weight Error (RUWE) is 2.842, while `good' solutions should have a RUWE near 1.0 (or < 1.4), and the ``Goodness of fit statistic of model wrt along-scan observations'' parameter is 38.3435, while good fits to the data should be $<3$.  The EDR3 parallax ($0.5051 \pm 0.0498$ mas) is also statistically incompatible with the DR2 value ($0.1530 \pm 0.0834$ mas).  The newer EDR3 value is likely more reliable (the older value has similar quality warnings), but we consider the possibility that there is a significant systematic error in the value. 
Relying on the spectroscopic stellar parameters and apparent magnitude, we can compare the Gaia distance with a luminosity-based distance. We use the previously calculated $\log (L/L_\sun)= 3.55 \pm 0.19$ to obtain the stellar bolometric magnitude. Adopting a solar bolometric luminosity $L_\sun = 3.828 \times 10^{26}\,$W and an absolute bolometric magnitude $M_\mathrm{bol} = 0\,$mag for a luminosity $L_{0} = 3.0128 \times 10^{28}\,$W (Resolution B2 of the XXIXth International Astronomical Union General Assembly in 2015), we obtain
\begin{equation}
M_{\rm bol}=-2.5\log{\frac{L}{L_{0}}} = -4.14 \pm 0.47.
\end{equation}

Applying a bolometric correction $BC_{\rm V} = -1.19 \pm 0.04$ \citep{2020MNRAS.495.2738P}, we find an absolute magnitude of $M_{\rm V} = -2.95 \pm 0.52\,$mag. The apparent magnitude is $m_V = 9.043 \pm 0.003\,$mag and the reddening $E(B-V) = 0.263 \pm 0.050$ \citep{TIC-8_2019}. Assuming $R(V)=3.1$, we find an extinction of $A_V= 0.817 \pm 0.156$.

Using the standard relation
\begin{equation}
M_V = m_V - 5\log d - 5 - A_V,
\end{equation}
where $d$ is in parsecs, we then find that \hd\ is  $d = 1721 \pm 543$ pc away, consistent within errorbars with the distance from Gaia\,EDR3 determined by \cite{Bailer-Jones2021}.

We further use the theoretical isochrones from \citet{Ekstrom_etal2012A&A...537A.146E} to determine an age $\log \tau \approx 7.68 \pm 0.10\,$yr. This is also illustrated in Fig.~\ref{fig:HRD}. Using the corresponding evolutionary tracks, we find that on the zero-age main sequence, \hd\ would have had  \teff\,$\approx 21\,000$ K and \logg $\,\sim 4.35$.

\subsection{Chemical abundances and stratification}
\label{sec:ResultsAbundances}

\begin{figure*}
    \centering
    \includegraphics[width=0.7\linewidth]{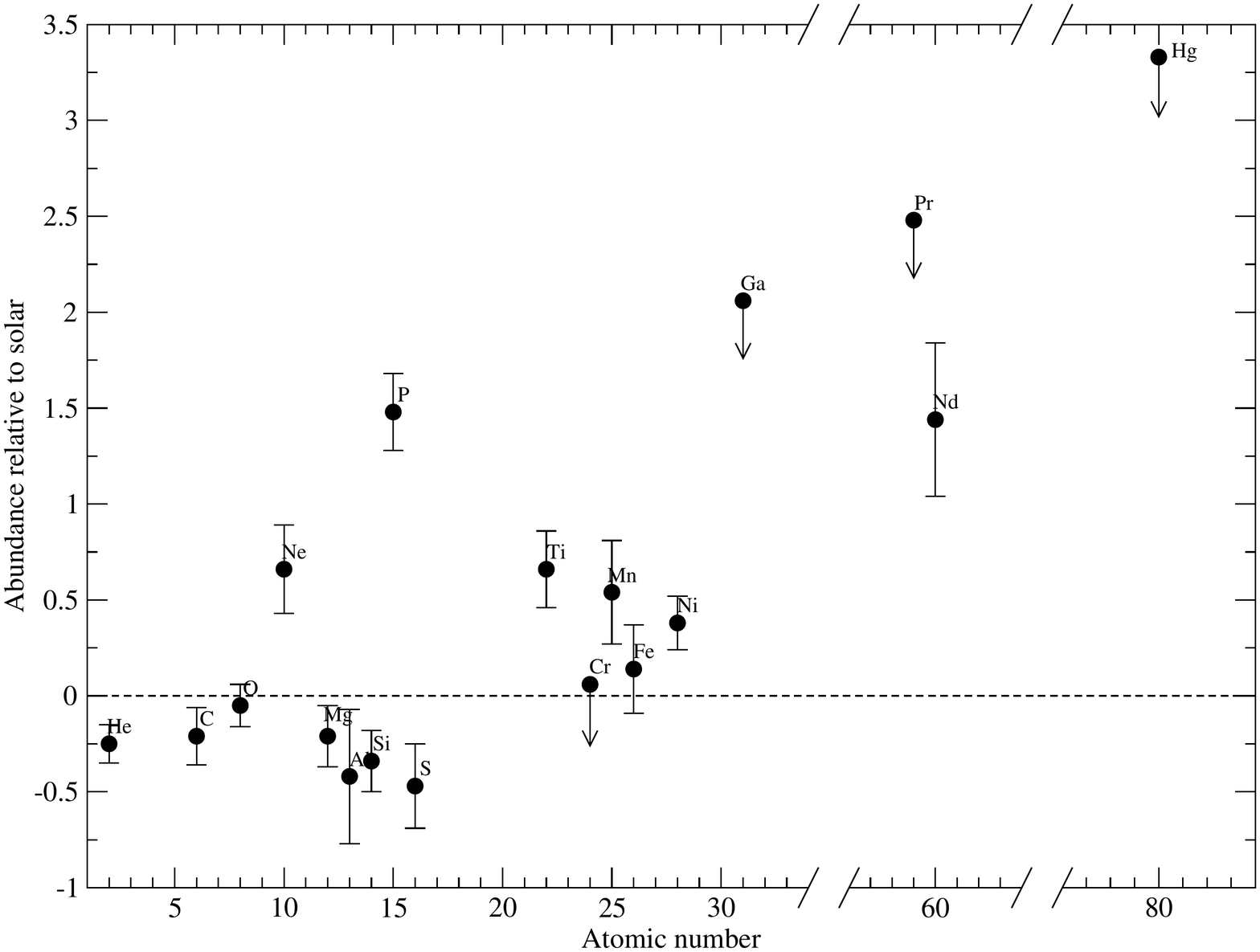}
    \caption{Derived abundances relative to the solar abundances of \citet{Asplund2009}.  Downward pointing arrows indicate upper limits only.  While many elements have near solar abundances, strong peculiarities are found for P, and moderate peculiarities for Ne and Ti.}
    \label{fig:abun-plot}
\end{figure*}

The best fit chemical abundances (see Table \ref{tab:abundances}) and plotted relative to solar abundances in Fig. \ref{fig:abun-plot} (in $\log(\mathrm{X}/\mathrm{H})$ units).  We find a strong overabundance of P by $1.48 \pm 0.20$ dex relative to solar, and significant overabundances of Ne by $0.65 \pm 0.23$ dex, and of Ti by $0.66 \pm 0.20$ dex.  Nd also appears to be strongly overabundant, although this is less certain as it is based on a few marginally detected lines.  Mn and Ni are marginally enhanced (by $0.54 \pm 0.27$ and $0.38 \pm 0.14$ dex, respectively).  C, O, Mg, Al, and Fe are all consistent with solar abundances (within $2\sigma$).  He appears to be weakly depleted ($-0.25 \pm 0.10$ dex), although this abundance is particularly sensitive to \teff.  Si also appears to be slightly depleted ($-0.34 \pm 0.16$ dex).  Thus \hd\ is clearly chemically peculiar, although it does not display the peculiarities typical of an HgMn star, and is only a very mild, marginally He-weak star (although it does have the enhanced P common to the He-weak PGa stars).  

The He abundance is particularly temperature sensitive, if our \teff\ were overestimated by $\sim$500 K the He abundance would be within uncertainty of solar (see Appendix \ref{AppendixTeff}).
Alternately, if our \teff\ was underestimated, that could lead to a larger He underabundance.  \ion{Ne}{I} lines may show some weak non-local thermodynamic equilibrium (NLTE) effects in B-type stars \citep{Alexeeva2020-NLTE-Ne-Bstars} (for \ion{Ne}{II} NLTE corrections appear to be negligible).  NLTE corrections for \ion{Ne}{I} at a \teff\ near 14\,000 K appear to reduce the Ne abundance by a few 0.1 dex \citep{Alexeeva2020-NLTE-Ne-Bstars}, possibly exceeding our error bar, but unlikely to produce an abundance consistent with solar, or chemically normal B-type stars.

\begin{figure}
    \centering
    \includegraphics[width=1.0\linewidth]{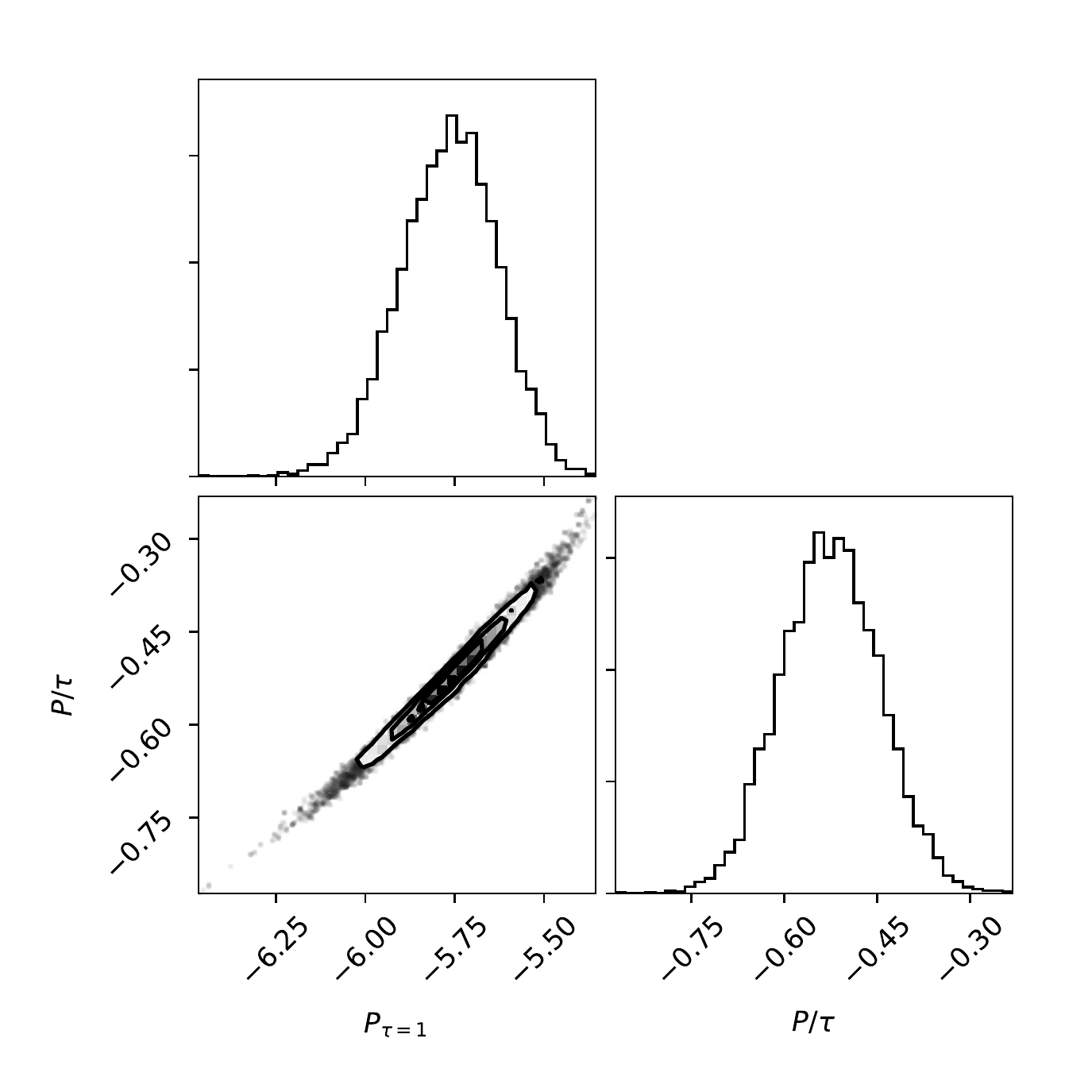}
    \caption{Posterior distributions of the model parameters for a linearly stratified P abundance from the MCMC analysis.  $P_{\tau = 1}$ represents the abundance at $\tau_{5000} = 1$, while $P/\tau$ is the slope in abundance with $\log \tau_{5000}$.  The lower left panel indicates strong covariance between these parameters.}
    \label{fig:model-P-params}
\end{figure}

From analyzing individual P lines (Sect.~\ref{sec:AnalysisStratification}) we find that lines formed mostly at smaller optical depths require significantly larger abundances of P to match the observations than lines formed deeper in the atmosphere.  The individual abundances and depths where $\tau_\ell = 1$ are plotted in Fig.~\ref{fig:strat-P-abun}, and we interpret this as evidence for stratification of P in the stellar atmosphere.  From simultaneously fitting P lines with synthetic spectra in an MCMC analysis, we find a clear negative slope in abundance with $\log \tau_{5000}$ ($-0.52 \pm 0.08$), and an abundance at $\tau_{5000} = 1$ that is still enhanced relative to solar ($-5.76^{+0.13}_{-0.14}$ dex), with an important covariance between these two parameters.  This P distribution is better constrained in the range of $\log \tau_{5000}$ between -1 and -2, and becomes poorly constrained above $\log \tau_{5000}$ -3 or below 0.  A set of model P distributions from the final Markov chain are presented in Fig.~\ref{fig:strat-P-abun} as overlapping gray lines, thus darker regions correspond to a higher density of models and a higher probability of the parameters.  The posterior probability densities of the parameters are shown in Fig.~\ref{fig:model-P-params}. From the combination of the single line analysis and the parametric MCMC model, we find good evidence for stratification of P, with larger abundances higher in the atmosphere.  However, since many lines of P are only marginally detected in our observations, higher precision data is needed before more details of the P distribution can be confidently constrained.

\subsection{Properties of the transiting secondary}\label{sec:secondary}

\begin{figure}[!ht]
\includegraphics[clip=,width=1.0\columnwidth]{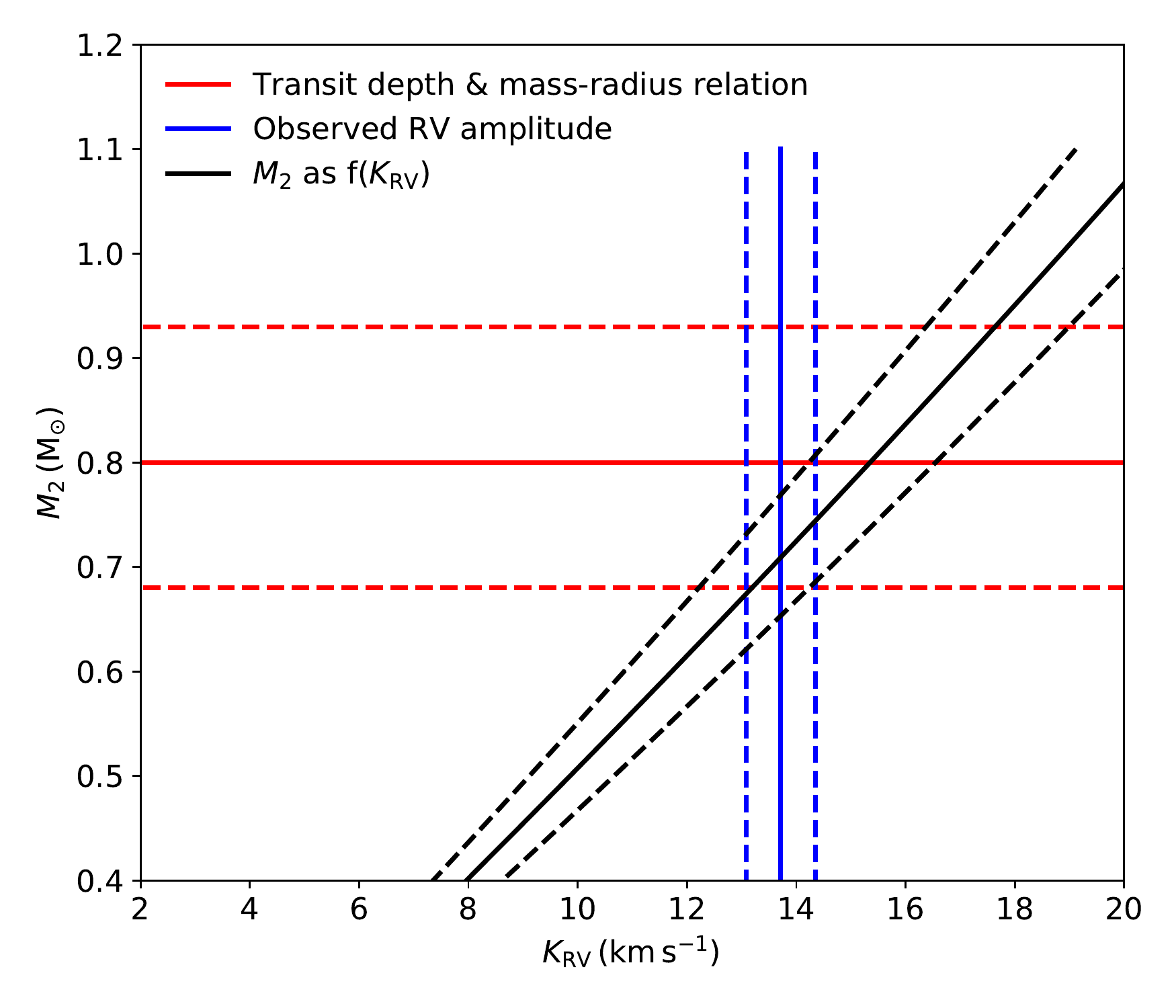}
\caption{The mass of the transiting secondary companion as a function of radial velocity perturbation on the primary. The mass of the secondary is constrained by the transit depth, using a main sequence stellar mass-radius relation (red lines), and by the radial velocity amplitude of the primary, assuming a circular Keplerian orbit (blue lines). The mathematical $K_{\rm RV}$--$M_{2}$ relation is also shown, with an uncertainty range reflecting the uncertain mass of the primary (black lines). }
\label{fig:secondary}
\end{figure}

The mass of the secondary star can be constrained in two ways. Firstly, using the observed transit depth (7.5 ppt) and our radius estimate for the primary, we obtain the radius of the secondary, $R_2 = 0.79^{+0.16}_{-0.13} R_\sun$.
In this radius estimate we assume the secondary is fully projected on the disk of the primary during photometric minimum, which is supported by the flat bottomed transits, and that the secondary contributes negligibly to the flux in the TESS band, which is supported by the small transit depth (suggesting a small cool object).  In that case the ratio $R_2 / R_1$ is the square root of the transit depth, and the uncertainty on $R_2$ is dominated by the uncertainty on $R_1$. 
Using a main sequence mass-radius relation for low-mass stars \citep[
${\leq}\, 1.5\,M_\sun$,][]{Eker_etal2018-relations}, we find a secondary mass $M_2 = 0.80^{+0.13}_{-0.12}\, M_\sun$. Secondly, we can obtain $M_2$ from the radial velocity perturbation amplitude of the primary, given by
\begin{equation}
K_{\rm RV} = \left( \frac{2\,\pi\,G}{P_{\rm orb}} \right)^{1/3} \times \left(\frac{M_{2}\,\sin i}{M_{1}+M_{2}}\right)^{2/3} \times (1-e^{2})^{-1/2}
\end{equation}
where $P_{\rm orb}$ is the orbital period, $e$ the orbital eccentricity, here assumed to be zero, $i$ is the inclination of the orbit, here assumed to be $90^\circ$, and $K_{\rm RV}=13.72\pm0.63\,$km\,s$^{-1}$ (Sect.\,\ref{sec:rvcurve}). 
This leads to a mass of $0.71^{+0.10}_{-0.09}$~\msun.
If we instead assumed $i = 80^\circ$, $M_2$ would increase by only $\sim$0.01 \msun.  If we assumed $e = 0.2$, that would decrease the mass by $\sim$0.02 \msun, and $e > 0.45$ would be needed to exceed our formal error bar. 
We summarize these constraints on $M_2$ in Fig.\,\ref{fig:secondary}. The results from both constraints are consistent, and point to $M_{2}$ in the range 0.7 -- 0.8 \msun.

\begin{table}[!ht]
\caption{Radial velocity measurements of \hd.}
\centering
\begin{tabular}{c c c c c c c}
\hline\hline
HJD & Phase & $RV$& $RV_{\rm err}$ \\
    &       &(${\rm km\,s^{-1}}$)& (${\rm km\,s^{-1}}$) \\
\hline
59112.24255 &0.5953 &0.7 &0.5 \\
59112.26930 &0.5964 &4.9 &0.5 \\
59112.32166 &0.5986 &7.2 &2.0 \\
59112.34425 &0.5995 &0.8 &1.8 \\
59112.36720 &0.6005 &7.2 &1.9 \\
59112.41234 &0.6023 &3.2 &1.5 \\
59112.43522 &0.6033 &3.0 &1.5 \\
59112.45896 &0.6043 &3.0 &1.5 \\
59112.50567 &0.6062 &-0.0 &1.9 \\
59112.52948 &0.6072 &-0.2 &3.0 \\
59174.46855 &0.1576 &-28.5 &1.9 \\
59174.49501 &0.1587 &-24.9 &2.6 \\
59177.15675 &0.2683 &-21.9 &1.7 \\
59177.17290 &0.2690 &-20.6 &1.6 \\
59177.19205 &0.2698 &-20.8 &1.5 \\
59177.20641 &0.2704 &-22.6 &1.4 \\
59296.54218 &0.1842 &-27.0 &1.9 \\
59296.56559 &0.1852 &-23.6 &1.7 \\
59296.58920 &0.1862 &-24.9 &1.9 \\
59306.42854 &0.5913 &-3.1 &2.7 \\
59306.45165 &0.5923 &-3.4 &2.6 \\
59306.47458 &0.5932 &-1.8 &2.3 \\
59311.48569 &0.7996 &-0.1 &3.4 \\
59311.50924 &0.8005 &0.4 &4.6 \\
59311.53243 &0.8015 &1.0 &4.5 \\
59315.46889 &0.9636 &-12.1 &4.9 \\
59315.49179 &0.9645 &-10.3 &4.4 \\
59315.51495 &0.9655 &-10.9 &3.7 \\
59323.39387 &0.2899 &-15.4 &2.1 \\
59323.41587 &0.2908 &-15.5 &2.0 \\
59323.44422 &0.2920 &-19.2 &1.9 \\
59323.47104 &0.2931 &-12.2 &2.3 \\
59323.50863 &0.2946 &-12.6 &2.3 \\
59323.53542 &0.2957 &-13.8 &2.3 \\
59324.38256 &0.3306 &-15.5 &3.7 \\
59324.40609 &0.3316 &-14.0 &3.7 \\
59324.42935 &0.3325 &-14.1 &3.4 \\
\hline
\end{tabular}
\tablefoot{Observations from the 1.5-m telescope at Tartu Observatory. HJD and phase correspond to midpoint of exposure.}
\label{tab:rv}
\end{table}

\begin{figure}[!htb]
    \centering
    \includegraphics[clip=,width=1.0\linewidth]{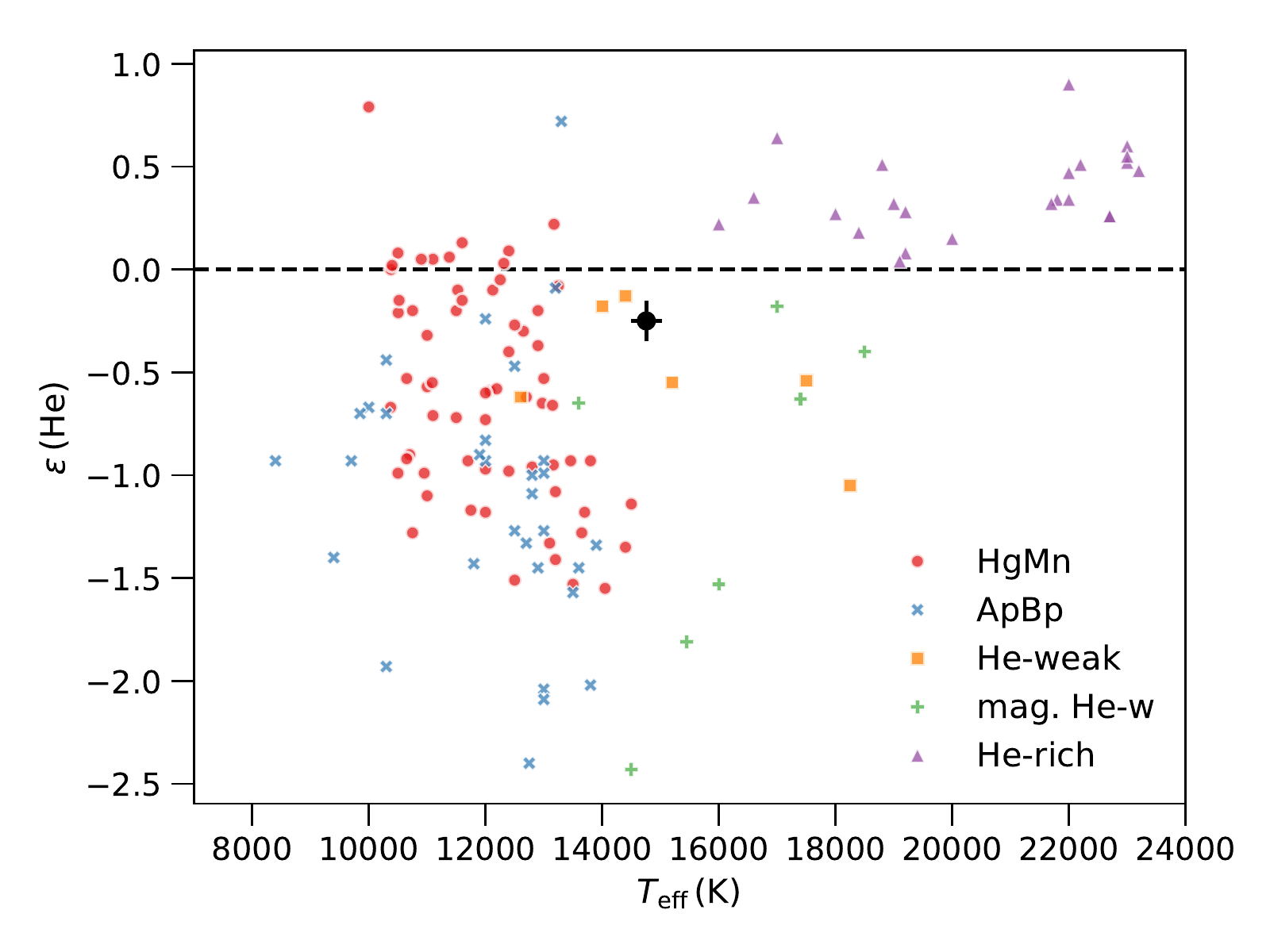}
    \includegraphics[clip=,width=1.0\linewidth]{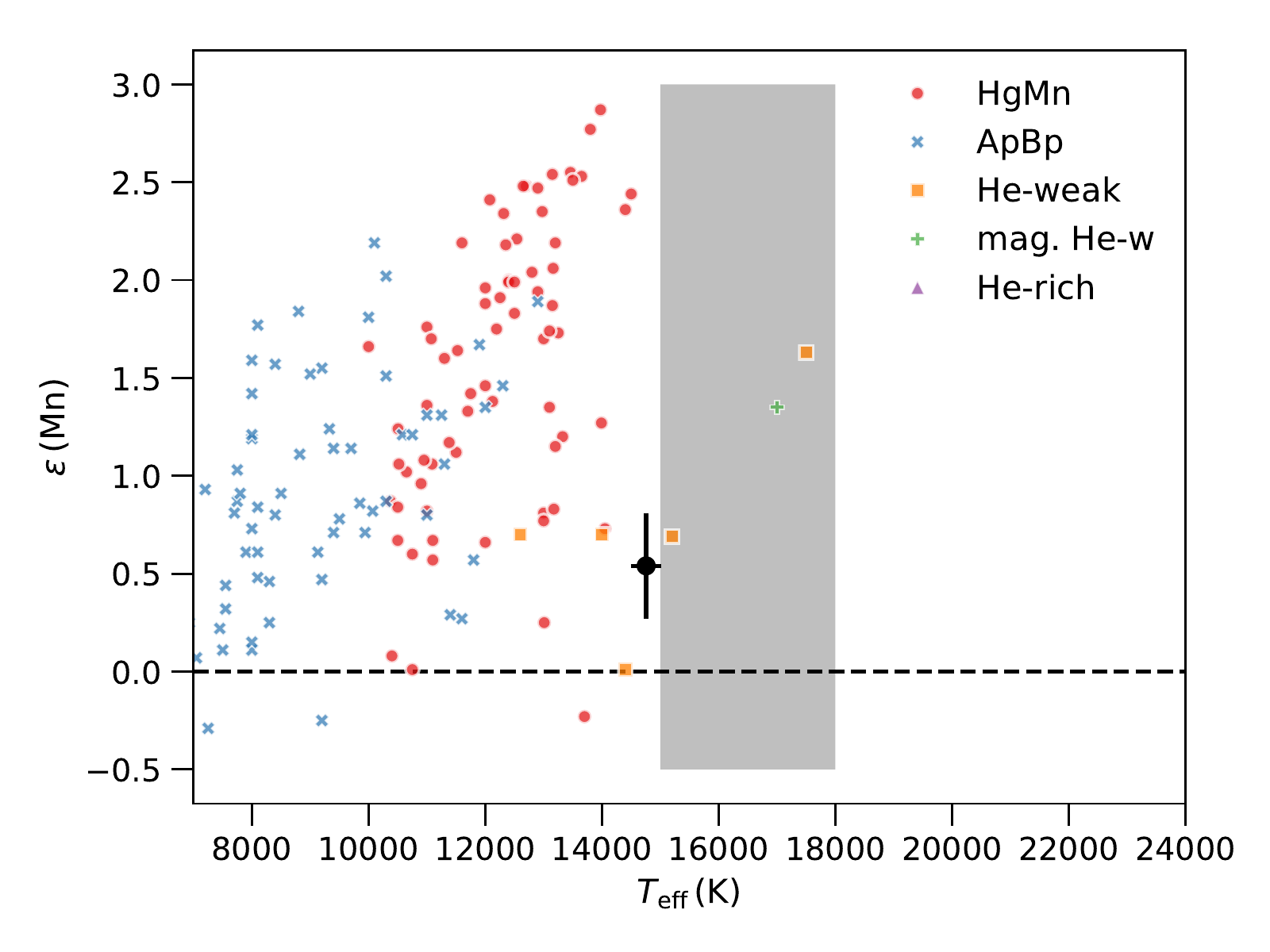}
    \includegraphics[clip=,width=1.0\linewidth]{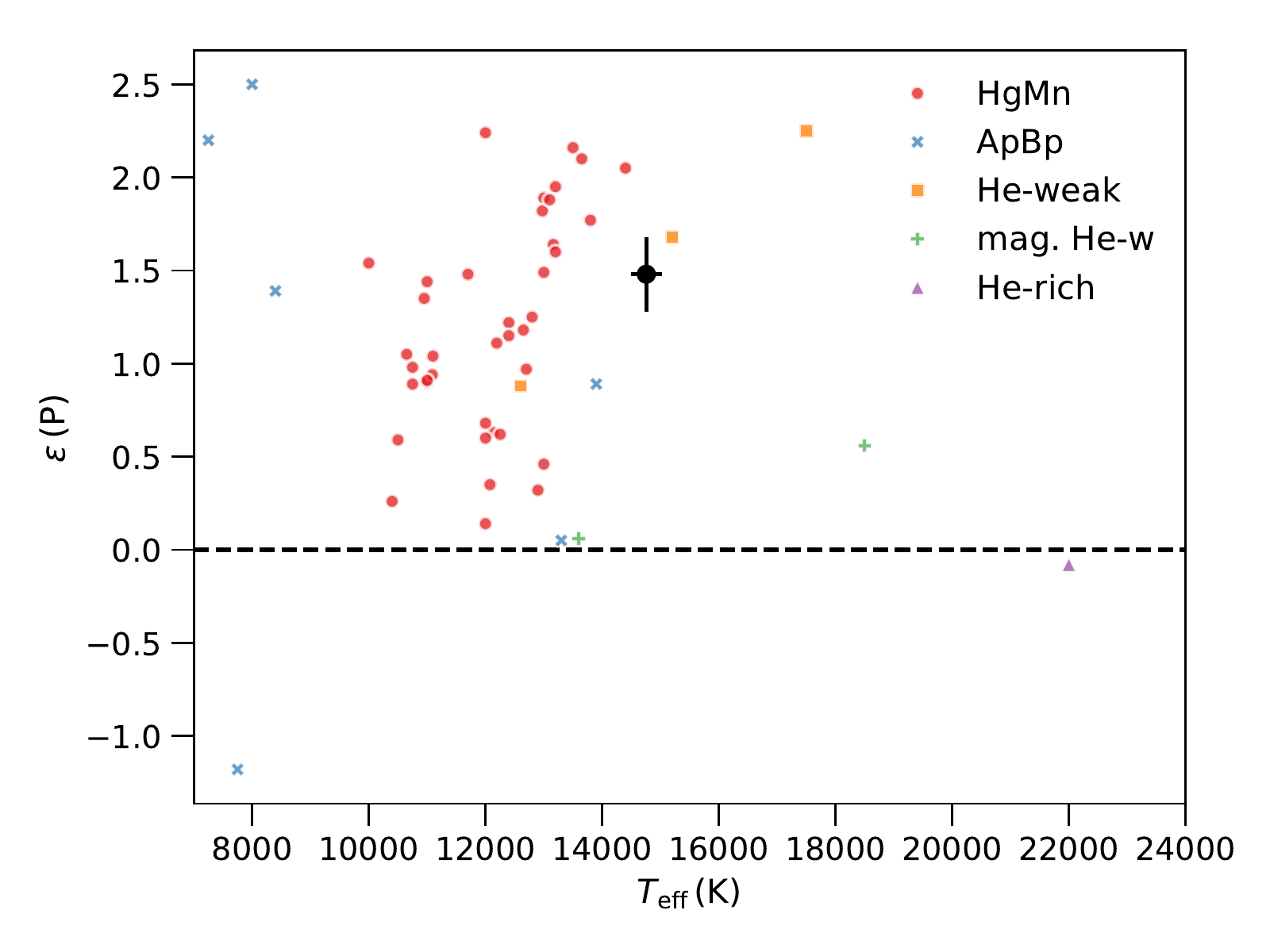}
    \caption{HD\,235349 in comparison to several major types of chemically peculiar early-type stars as a function of stellar $T_{\rm eff}$. \emph{Top panel: }The helium abundance in \hd\ relative to the sun, with the comparison sample from \citet{Ghazaryan2018-catalogue-Ap-Am-HgMn} and \citet{Ghazaryan2019-HeWeak-HeStrong-roAp} plotted in color. \emph{Middle} and \emph{bottom panels: }Same as above, but for manganese and phosphorus. The gray region in the middle panel indicates the $T_{\rm eff}$ range where the Mn enhancement should drop off toward higher temperatures \citep{Alecian-Michaud1981}.}

    \label{fig:sample}
\end{figure}

\section{Discussion}\label{sec:discussion}

\subsection{Spectral type and effective temperature}

Using a Balmer line analysis, we determined an effective temperature \teff\,$=14\,757 \pm 264\,$K and \logg\,$= 3.34 \pm 0.11$ for \hd.  This yields a spectral type B6 III
\citep{PecautMamajek2013}. These results are consistent with the highest \teff\ and earliest spectral types from the wide range of past estimates. As noted earlier, the ExoFOP database lists two conflicting effective temperature values for \hd. The latest value, from the TESS project, is \teff\,$=8381\pm409\,$K. In the databases listed on ViZieR\footnote{\texttt{http://simbad.u-strasbg.fr/simbad/sim-fid}}, a wider range of $T_{\rm eff}$ estimates can be found, corresponding to a spectral type range from mid-A to mid-B. Spectral type classifications spanning nearly 100 years have typically been B8 or B5 \citep{Cannon1927, HillLynasGray1977}.

\subsection{Chemical peculiarity}

The composition of \hd\ is clearly peculiar, with a strong overabundance of phosphorus (P, $+1.48$ dex), neon (Ne, +0.65~dex),  and neodymium (Nd, +1.44~dex). The upper limits for gallium and praseodymium also allow for enhancements of up to 2\,dex. Furthermore, Ti, Mn, and Ni may be slightly (a factor of $0.3$ to $0.6\,$dex) enhanced, while Si, and S appear underabundant by a similar factor and He appears to be weakly underabundant at 0.25 dex (although the He abundance is sensitive to \teff).

The star is not an obvious member of the mercury-manganese (HgMn) group of chemically peculiar stars, which extends up to temperatures similar to \hd\ \citep[up to $\sim$14\,000~K, e.g.][]{Ghazaryan2018-catalogue-Ap-Am-HgMn}. Mn is not strongly overabundant in \hd, and Hg (particularly the \ion{Hg}{II} 3984 \AA\ line) is not detected (see Fig.\,\ref{fig:spectrumHg}; although the upper limit is only $<3.3$ dex enhanced).  Thus the star cannot be confidently classified as an HgMn star.  
We compare chemical abundances for a few elements of interest, for different samples of peculiar stars\footnote{
We have divided the sample of helium weak stars from \citet{Ghazaryan2019-HeWeak-HeStrong-roAp} into those with good evidence for strong magnetic fields in the literature (`mag.\ He-w') and those lacking evidence for a magnetic field (`He weak'), since the presence of a strong magnetic field modifies atomic diffusion. }, as a function of \teff\ in Fig.~\ref{fig:sample}.
The strongest similarity between \hd\ and HgMn-type stars is that HgMn stars are typically overabundant in P.
\citet{SmithDworetsky1993} and \citet{Ghazaryan2018-catalogue-Ap-Am-HgMn} note a population of hot (>13\,000 K) HgMn stars with mild Mn overabundances (Fig.~\ref{fig:sample}), which is closer to \hd.

While the He abundance is only mildly below solar, the P- and potentially Ga-rich pattern is consistent with the rare class of He-weak PGa chemically peculiar stars. These are weakly- or non-magnetic stars which may represent the high temperature end of HgMn stars \citep{Borraetal1983}. 
There are a few stars that have been classified as He-weak, despite apparently very modest He underabundances \citep[e.g.][see Fig.~\ref{fig:sample}]{Glagolevskij-HeWeak-marginal}. Thus \hd\ has similarities to the He-weak PGa stars, although given the mild He underabundance (and sensitivity of the He abundance to \teff) we do not give it this classification with a high level of confidence.  

We propose that \hd\ represents an intermediate case between HgMn and He-weak PGa stars, perhaps with some chemical peculiarities weakened by mixing from the relatively rapid rotation of the star.  There is significant evidence to support the idea that He-weak PGa stars represent a continuation of HgMn stars to higher temperatures \citep[e.g.][]{Borraetal1983, Rachkovskaya2006-HeWeakPGa, Hubrig2014-HeWeakPGa, Monier2021-Maia-HeWeakPGa}, with the differences in abundances due to differences in atomic diffusion with increasing temperature and perhaps an increasingly important stellar wind.  \hd\ sits near the border in \teff\ between these classes of objects \citep{Ghazaryan2018-catalogue-Ap-Am-HgMn, Ghazaryan2019-HeWeak-HeStrong-roAp}, supporting the idea that it is intermediate between these classes.

Theoretical atomic diffusion calculations support this general picture.  Early work by \citet{Alecian-Michaud1981} on diffusion suggested that the overabundance of Mn in HgMn stars 
should fall off with increasing \teff\ somewhere in the \teff\ $\sim 15000$ to 18000 K range.
\hd\ falls on the edge of this range, and this trend is generally consistent with other observations of Mn abundances (see Fig.\,\ref{fig:sample}).  More recent work by \citet{Alecian2019-Alecian-Stift-diffusion} consider time dependent diffusion including mass-loss, with a focus on HgMn stars.  They consider stratification of P, finding overabundances higher in the atmosphere, and finding that these overabundances can be greatly enhanced by a weak magnetic field.  

We find clear evidence for vertical stratification in the abundance of P, with larger abundances higher in the atmosphere.  This can be adequately modeled with a linear distribution of P in $\log \tau_{5000}$, although higher S/N observations are needed to provide better constraints on the vertical distribution of P. Stratification of P, as well as some other elements, has been detected in a few HgMn stars (in HD~53929 and HD~63975 by \citealt{Ndiaye2018-HgM}, and in HD~161660 by \citealt{Catanzaro2020-HgM-strat}), and also in HD~213781 which may be a blue horizontal branch star or an evolved HgMn star \citep{Kafando2016-BHB-abun-strat}.  We find qualitatively consistent results with an apparently gradual increase in P abundance to shallower optical depths.  This is also qualitatively consistent with the atomic diffusion models of P by \citet{Alecian2019-Alecian-Stift-diffusion}.

\hd\ has a higher \vsini\ than is typical for an HgMn star or a He-weak PGa star.  However, we are likely seeing the full equatorial rotation velocity (with $i$ close to $90^\circ$), since the system is an eclipsing binary, and with the relatively short period the orbital and rotational axes are likely aligned.
The recent survey of HgMn stars by \citet{Chojnowski2020-HgMn-SDSS-survey} found an average \vsini\ of 28 \kms, but also 13 stars with \vsini\ above 60 \kms.  \citet{Gonzalez2021-HgMn-high-vsini} report a HgMn star with \vsini\ $= 124$ \kms\ as the highest \vsini\ HgMn star yet discovered.  
The high \vsini\ of \hd\ likely enhances mixing in the star through meridional circulation, which may reduce the amount of chemical stratification and the strength of the observed peculiarities of some elements.  

We find a low \logg~$=3.34 \pm 0.11$ for \hd, which implies the star has likely evolved off the main sequence.  However, that \logg\ is not unprecedented among chemically peculiar stars in this temperature range, e.g.\ the HgMn star HD 63975 \citep[\logg\ = 3.27,][]{Ndiaye2018-HgM},
the He-weak PGa star HD 23408 \citep[\logg\ =  3.3,][]{Mon1981-HeWeak, Monier2021-Maia-HeWeakPGa}, and the He-weak (probably not of the PGa type) HD 5737 \citep[\logg\ =  3.2,][]{Leone1997-HeWeak-Teff-Corr, Saffe2014-HeWeak}.  The low \logg\ may influence atomic diffusion, and contribute to the atypical pattern of chemical abundances observed, although the impact of \logg\ on atomic diffusion is not well studied.

\subsection{Nature of the transiting object}

The large radius of \hd\ rules out a planetary explanation for the transits in the TESS lightcurve, instead strongly favoring a slightly sub-solar mass main sequence star. 
Recently, it was found from speckle interferometry that \hd\ has a close fainter ($\Delta m=4.1^{m}$ at 832 nm) companion at the angular distance of 0.263 arcseconds \citep{Howell_2021}. It is probable that \hd\ and this close companion form a true bound stellar system, the distance determined in current work implies the separation between two stars is at least $452 \pm 143$~AU (for a distance of $1721 \pm 543$ pc), but this is too distant to be the transiting object.
We rule out contamination of the photometry from a different source blended within a TESS pixel (e.g.\ an eclipsing binary projected near \hd). There is no sign of another bright star within a $21''$ diameter area, and our optical spectra also contain no indications of another blended bright star. If the ${\sim}0.8\,$\% dip in the summed lightcurve were from a different object it would require a faint binary where one component is completely covered, but this would produce a triangular transit whereas the TESS data clearly show a flat bottom. We conclude that the transiting companion must be bound to \hd, this assumption leads to a companion mass ${\approx}0.75\,$\msun (Sect.\,\ref{sec:secondary}).

There are very few known eclipsing binaries among HgMn and He-weak stars, although new data from TESS is helping to change this.  Binarity is often thought to be important for generating these chemically peculiar stars, by slowing rotation rates through tidal interactions, thereby allowing atomic diffusion to proceed efficiently.  Despite this there are only ten known eclipsing binaries among HgMn stars:
AR Aur \citep[HD 34364,][]{Nordstrom1994-ARAur-bin, Hubrig2006-ARAur-Var, Folsom2010-HgMn-bin}, TYC 455-791-1 \citep{Strassmeier2017-HgMn-bin}, V772 Cas \citep[HD 10260,][]{Kochukhov2021-HgMn-bin}, V680 Mon \citep[HD 267564,][]{Paunzen2021-HgMn-bin}, HD 72208 \citep{Wraight2011-STEREO-eclipse, Kochukhov2021-HgMn-eclipse-tess}, and TYC 4047-570-1, HD 36892, HD 50984, HD 55776, HD 53004 \citep{Kochukhov2021-HgMn-eclipse-tess}. 
Possible eclipses have been reported in HD 161701 \citep[][see also \citealt{Gonzalez2014-HD161701-HgMn-bin}]{Wraight2011-STEREO-eclipse}, and single eclipse-like events in TESS data have been reported for HD 34923 and HD 99803 \citep{Kochukhov2021-HgMn-otherVar-tess}, but more observations are needed to confirm these.
The eclipsing binary HD 66051 is of interest \citep{Niemczura2017-maybeHgMn} although it appears to contain a magnetic Bp star \citep{Kochukhov2018-HD66051-Bp}, and HD 62658 was also recently discovered to contain an eclipsing magnetic Bp star \citep{Shultz2019-HD62658-Bp}.
We are not aware of any eclipsing binaries among the He-weak PGa stars, perhaps due to their rarity and the difficulty in firmly establishing this classification.  
In this context, the eclipsing binary nature of \hd\ is an important addition to the known hot chemically peculiar stars.

\section{Conclusions}

   \begin{enumerate}
      \item Transit-like events visible in the TESS lightcurve are in good agreement with eclipses by a low-mass stellar companion in a binary system, not an exoplanet.  The transit depth and radius of the primary star imply a stellar radius for the transiting object, of $0.79^{+0.16}_{-0.13} R_\sun$.
      \item Radial velocity variability was detected in the observed spectra, which varies in phase with the transit events, implying a stellar mass for the transiting object.  This dynamical mass of $0.71^{+0.10}_{-0.09}\,$\msun\ is consistent with the mass implied by the inferred radius of the transiting object. 
      \item Based on fitting observations with model spectra, we classify the primary 
      as a B6 III star, with \teff\ $=14\,757 \pm 264$ K and \logg\ $=3.34 \pm 0.11$.
      \item The primary of the system is a chemically peculiar star, showing a large overabundance of phosphorus, overabundances of neon and titanium, and a weak underabundance of helium. 
      Clear evidence for stratification of phosphorus in the atmosphere of the star was found, with larger abundances at shallower optical depths.
      The spectral line of \ion{Hg}{II}\, at 3984.0~\AA~was not detected, suggesting this is not a typical HgMn star. Instead we propose this may be an He-weak PGa star or an intermediate between the He-weak PGa stars and the HgMn stars.
      \item This appears to be the first star discovered with He-weak PGa characteristics in an eclipsing binary system.
   \end{enumerate}

\begin{acknowledgements}
The authors gratefully acknowledge financial support from the Estonian Ministry of Education and Research through the Estonian Research Council institutional research funding IUT40-1 and from the European Union European Regional Development Fund project KOMEET 2014-2020.4.01.16-0029. This work is based on observations collected with the 1.5-m telescope at Tartu Observatory, Estonia, and includes data collected by the TESS mission. Funding for the TESS mission is provided by the NASA’s Science Mission Directorate. This work has also made use of the VALD database, operated at Uppsala University, the Institute of Astronomy RAS in Moscow, and the University of Vienna; the VizieR catalogue access tool CDS, Strasbourg, France (DOI : 10.26093/cds/vizier); and the Exoplanet Follow-up Observation Program website, which is operated by the California Institute of Technology, under contract with the National Aeronautics and Space Administration under the Exoplanet Exploration
Program.
 \end{acknowledgements}

\bibliographystyle{aa}
\bibliography{HD235349}

\clearpage

\begin{appendix}

\section{Instrument parameters}\label{AppendixA}
The key parameters of the AZT-12 telescope at Tartu Observatory, and the long-slit spectrograph mounted there, are given in Table~\ref{tab:AZT12}.

\begin{table}[h]
\caption{Parameters of the telescope AZT-12 and the spectrograph ASP-32 at Tartu Observatory.}
\centering
\begin{tabular}{l c}
\hline\hline
Parameter & Specification \\
\hline
Telescope Type & Cassegrain reflector\\
Primary Diameter & 1.50 m\\
Secondary Diameter & 0.36 m\\
Focal Ratio &F/16\\
Cassegrain Focal Length & 24.0 m\\
Cassegrain focus plate scale & 116 ${\rm \mu m\,arcsec^{-1}}$\\
\hline
Spectrograph type & Long-slit spectrograph \\
                  & at Cassegrain focus \\
Spectrograph slit width & 2.2 arcsec \\
Spectrograph slit length & $\leq1$ arcmin \\
Gratings (${\rm l\,mm^{-1}})$ & 300, 600, 1200, 1800, 2400 \\
Spectral resolution & $1000\ldots11000$ \\
\hline
\end{tabular}
\label{tab:AZT12}
\end{table}

\section{Additional temperature constraints}
\label{AppendixTeff}

\begin{figure*}[htb]
    \centering
    \includegraphics[width=0.4\linewidth]{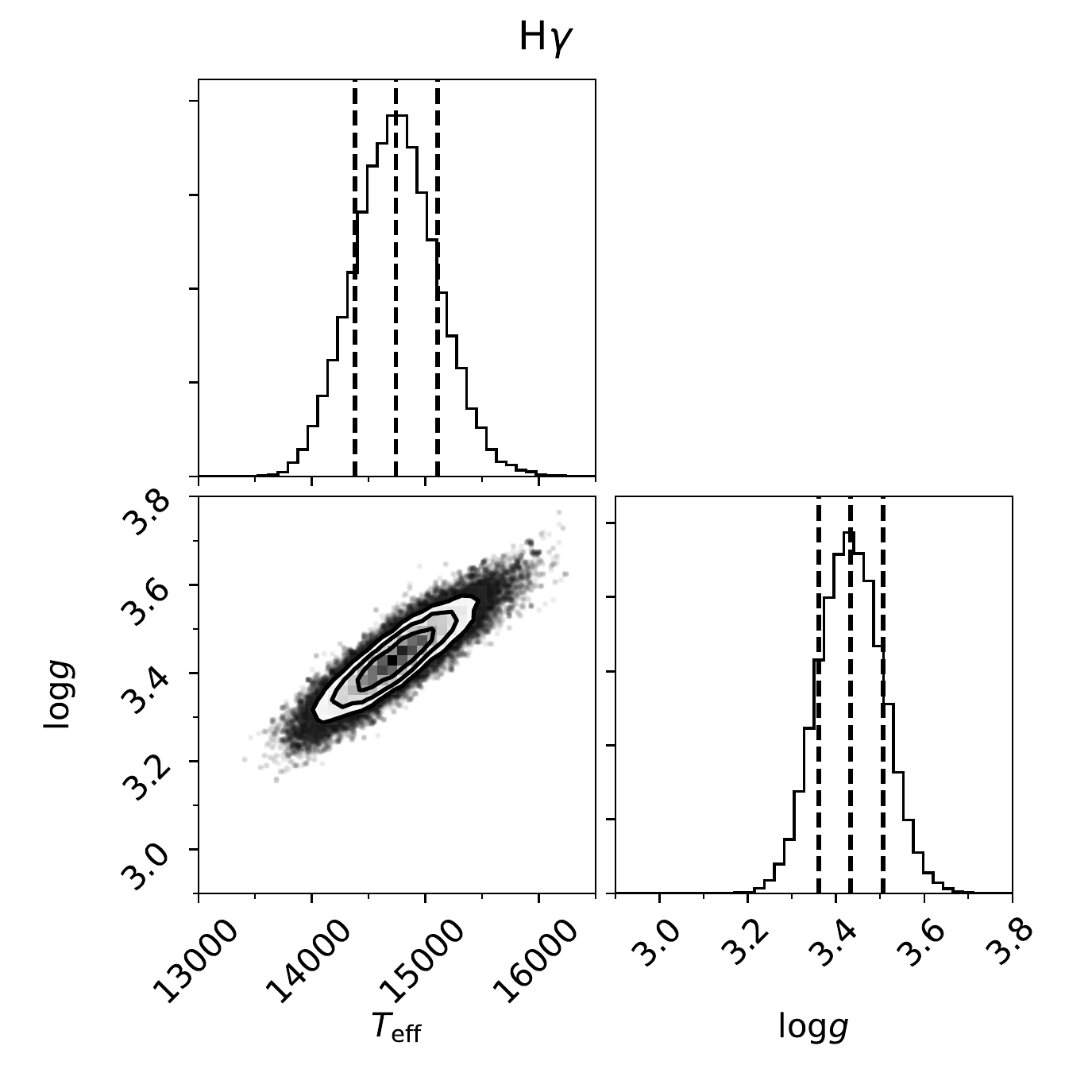}
    \includegraphics[width=0.4\linewidth]{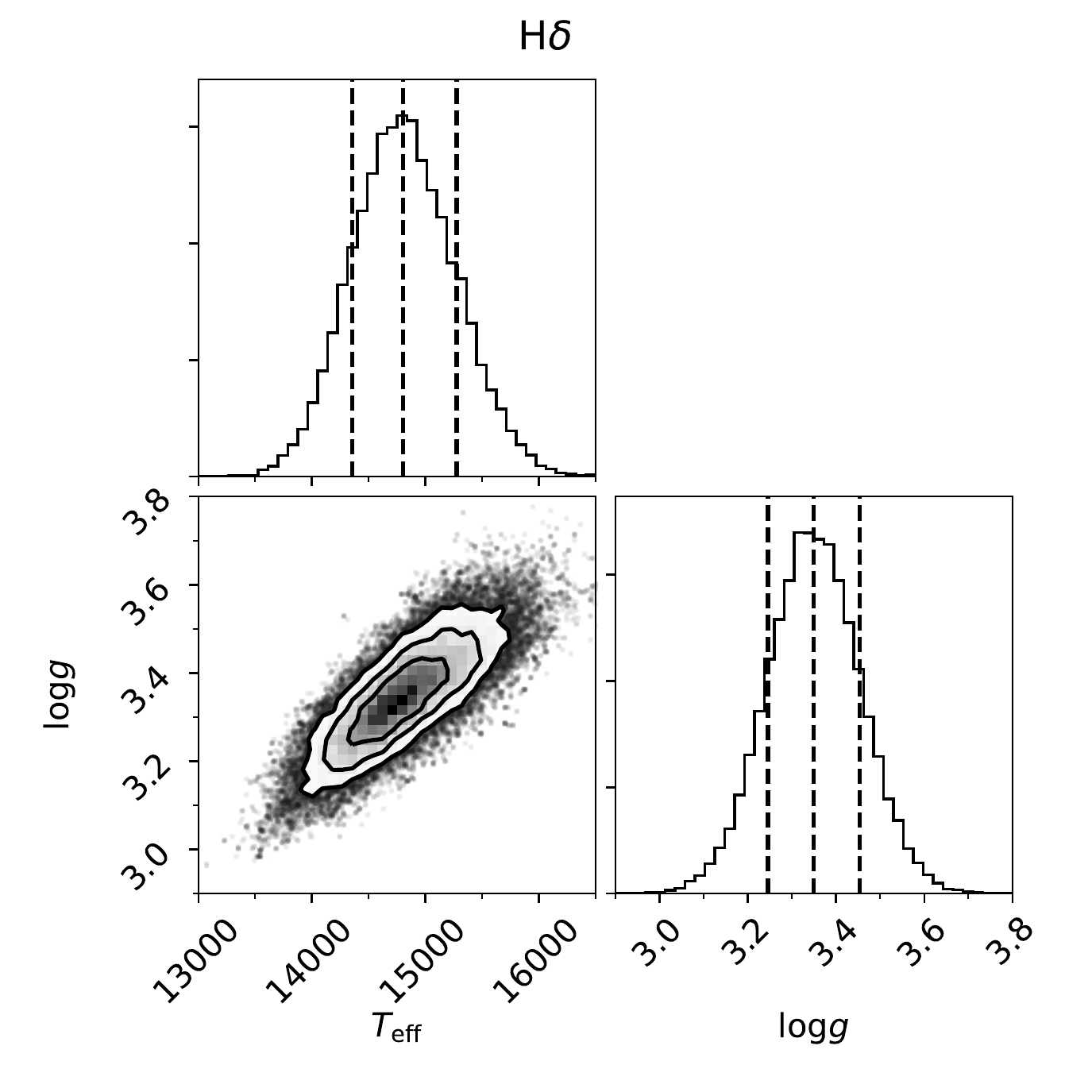}
    \includegraphics[width=0.4\linewidth]{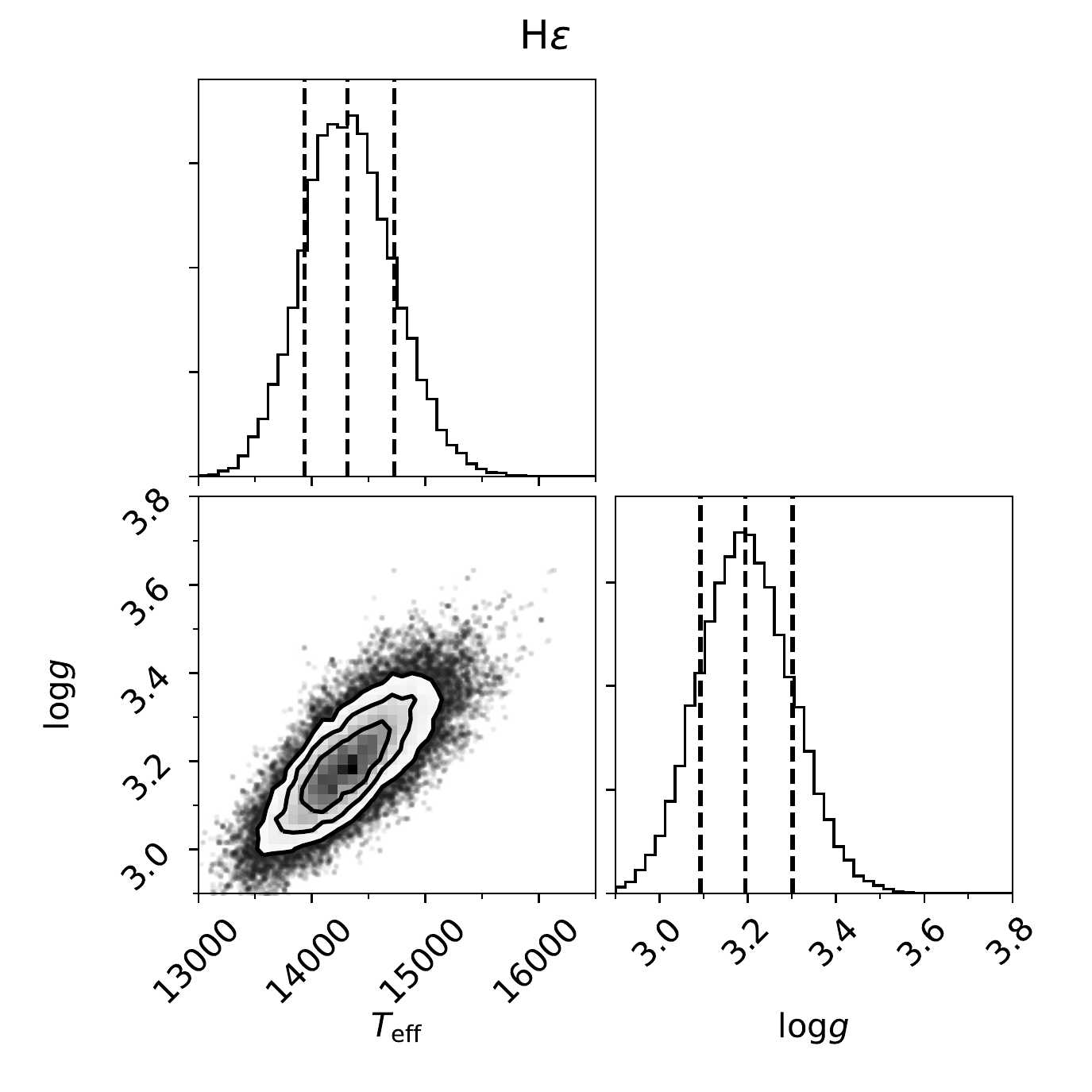}
    \includegraphics[width=0.4\linewidth]{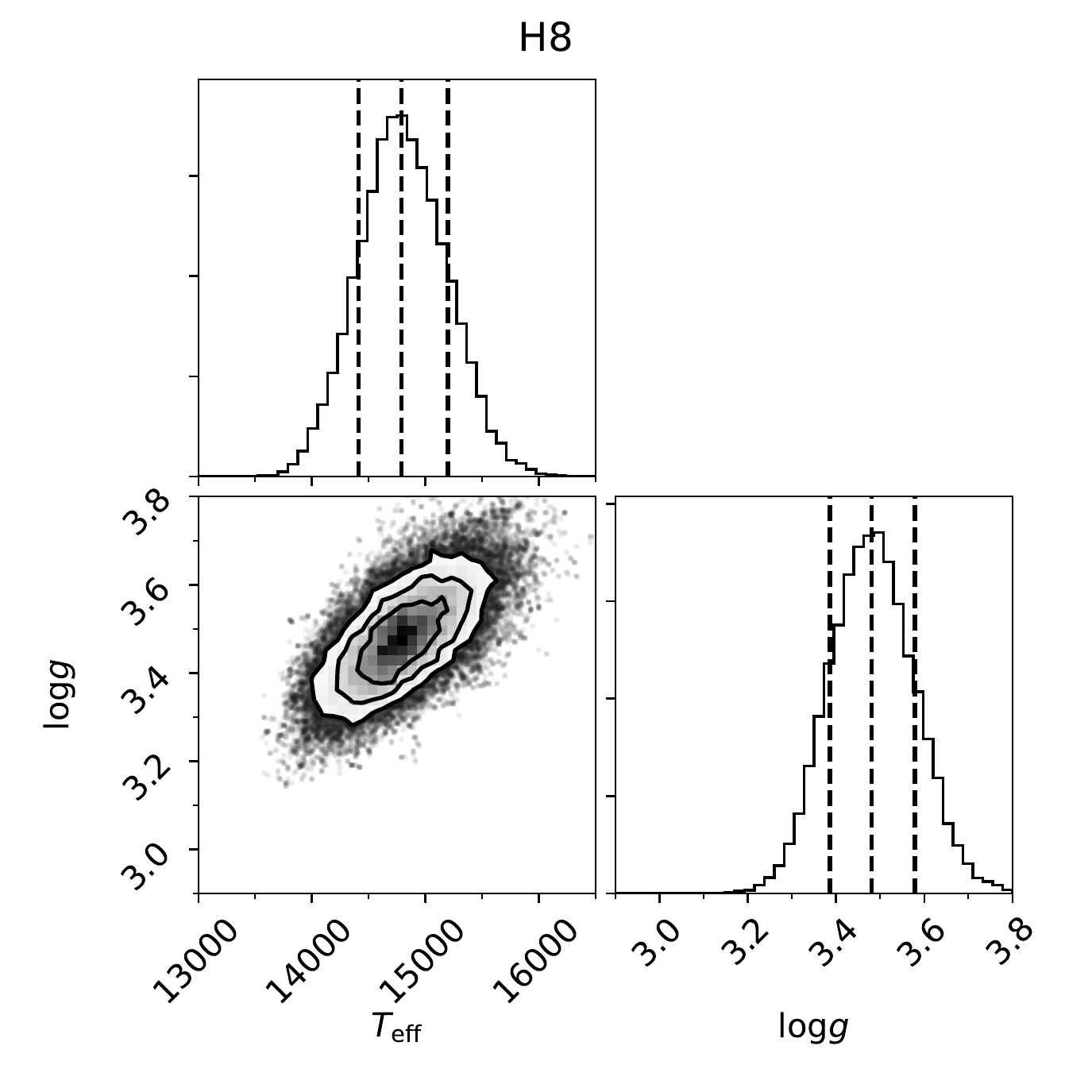}
    \includegraphics[width=0.4\linewidth]{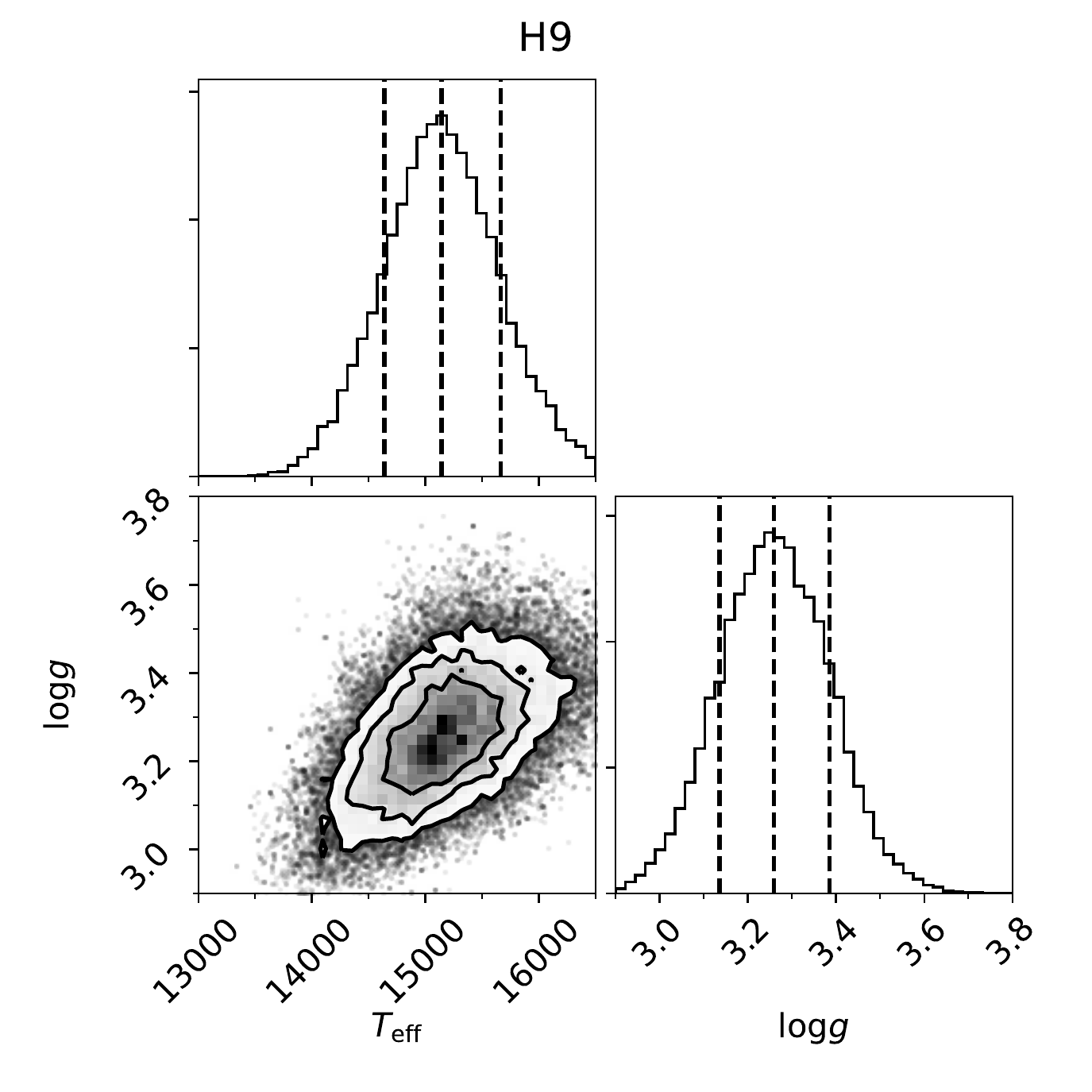}
    \caption{Posterior probability distributions of \teff\ and \logg\ from the MCMC analysis of Balmer lines, after marginalizing over the continuum parameters (using corner.py, \citealt{ForemanMackey_corner}).  Plots are shown for the H$\gamma$, H$\delta$, H$\epsilon$, H8, and H9 lines. Vertical dashed lines indicate the median and $\pm 1\sigma$, contours indicate 1, 1.5, and $2\sigma$ levels.
    }
    \label{fig:balmerLine-cornerPlots}
\end{figure*}

The primary constraints on \teff\ and \logg\ are from the MCMC based fit to five Balmer lines, as described in Sect.~\ref{sec:BalmerLineAnalysis}.  Here we include plots of the posterior distributions (marginalized over the three continuum polynomial coefficients) from the MCMC analysis for each individual line in Fig.~\ref{fig:balmerLine-cornerPlots}.  We find an important covariance between \teff\ and \logg\ for all lines.  Covariances between \teff\ or \logg\ and the continuum parameters (not shown) are generally weak.

As a secondary analysis we derive \teff\ and \logg\ from the fit to metallic lines.  Since the star is chemically peculiar, these parameters must be determined simultaneously with chemical abundances, when using metal lines.  This was done using the same analysis methodology as described in Sect.~\ref{sec:MetalLineAnalysis} for the metal and He lines, using the same spectral windows and line data, but including \teff\ and \logg\ as additional free parameters.  The resulting fits to the observations are of a comparable quality to the previous metal line analysis.  The best fit values averaged over spectral windows are in Table \ref{tab:alternate-abundances}, with uncertainties taken as the standard deviation or estimated as before for elements with abundances from less than three windows.  This analysis provides a weaker constraint on \teff\ and \logg\ than the Balmer line analysis, but the results are consistent within uncertainties.  Similarly, the chemical abundances are consistent within uncertainty, but may have slightly larger systematic errors due to the larger uncertainties in \teff\ and \logg.  The most important difference in abundance is that He becomes more abundant and uncertain (with a larger scatter between windows), putting it within $1\sigma$ of the solar abundance.  Thus the lower \teff\ from the metal line analysis suggests the star may not be He weak, but it also produces less consistent results from different He lines, suggesting it may be less reliable.  

\begin{table}[!ht]
    \centering
    \caption{Alternate stellar parameters and abundances, based solely on metallic and He lines.} 
    \begin{tabular}{lccl}
\hline\hline
Parameter & Value & Solar & \#  \\  
\hline
\teff\ (K)     & $14\,189 \pm 492 $ \\
\logg          & $3.43  \pm 0.21$ \\
\vsini\ (\kms) & $64.8 \pm 7.1$ \\
$v_{\rm mic}$ (\kms) & $2.19  \pm 0.83$ \\
\hline
He   & $-1.11 \pm 0.18$ & -1.07  & 3 \\
C    & $-3.61 \pm 0.15$ & -3.57  & 1 \\
O    & $-3.40 \pm 0.11$ & -3.31  & 2 \\
Ne   & $-3.30 \pm 0.23$ & -4.07  & 2 \\
Mg   & $-4.63 \pm 0.14$ & -4.40  & 4 \\
Al   & $-5.96 \pm 0.37$ & -5.55  & 3 \\
Si   & $-4.79 \pm 0.15$ & -4.49  & 5 \\
P    & $-5.02 \pm 0.24$ & -6.59  & 5 \\
S    & $-5.20 \pm 0.21$ & -4.88  & 4 \\
Ti   & $-6.53 \pm 0.20$ & -7.05  & 2 \\
Cr   & $< -6.3        $ & -6.36  & 1 \\
Mn   & $-6.21 \pm 0.26$ & -6.57  & 2 \\
Fe   & $-4.47 \pm 0.21$ & -4.50  & 4 \\
Ni   & $-5.51 \pm 0.14$ & -5.78  & 2 \\
Ga   & $< -6.9        $ & -8.96  & 1 \\
Pr   & $< -8.8        $ & -11.28 & 1 \\
Nd   & $-9.13 \pm 0.40$ & -10.58 & 1 \\
Hg   & $< -7.5        $ & -10.83 & 1 \\
\hline\hline
    \end{tabular}
    \tablefoot{Solar abundances from \citet{Asplund2009} are listed for comparison. The number of individual spectral windows used to constrain a given element is indicated in the last column.  These results are less precise, but consistent with, the results using the Balmer lines for \teff\ and \logg\ (Table \ref{tab:abundances}). }
    \label{tab:alternate-abundances}
\end{table}

\end{appendix}

\clearpage

\end{document}